\documentclass[fleqn,usenatbib]{mnras}

\usepackage{newtxtext,newtxmath}
\usepackage{amsmath,bm}	
\usepackage[T1]{fontenc}
\usepackage{xspace}
\DeclareRobustCommand{\VAN}[3]{#2}
\let\VANthebibliography\thebibliography
\def\thebibliography{\DeclareRobustCommand{\VAN}[3]{##3}\VANthebibliography}

\usepackage{graphicx}	
\usepackage{amsmath}	

\def\({\,\left(}
\def\){\,\right)}
\def\[{\left[}
\def\]{\right]}

\def\mhmpc{\,h^{-1}{\rm Mpc}}

\def\mhgpcc{\,h^{-3}{\rm Gpc^3}}

\def\mhmsun{\,h^{-1}{\rm M_{\odot}}}

\newcommand{\vmult}{\,\rm v_{mult}\xspace}
\newcommand{\vmean}{\,\rm v_{mean}\xspace}
\newcommand{\vgeom}{\,\rm v_{geom}\xspace}
\newcommand{\vmatch}{\,\rm v_{match}\xspace}
\newcommand{\vmeanresc}{\,\rm v_{mean}^{resc}\xspace}
\newcommand{\vmultresc}{\,\rm v_{mult}^{resc}\xspace}

\newcommand{\DD}{\,\mathrm{DD}\xspace}

\newcommand{\DR}{\,\mathrm{DR}\xspace}
\newcommand{\RR}{\,\mathrm{RR}\xspace}

\newcommand{\nsv}{\,n_{\rm sv}\xspace}
\newcommand{\nb}{\,n_{\rm b}\xspace}
\newcommand{\ns}{\,n_{\rm s}\xspace}

\newcommand{\nbs}{\,n_{\rm bs}\xspace}

\newcommand{\rr}{\mathbf{r}}

\title[Covariance matrix]{Creating Jackknife and Bootstrap estimates of the covariance matrix for the two-point correlation function}

\author[Faizan G. Mohammad et al.]{
Faizan G. Mohammad,$^{1,2}$\thanks{E-mail: faizan.mohammad@uwaterloo.ca}
Will J. Percival,$^{1,2,3}$
\\
$^{1}$Waterloo Center for Astrophysics, University of Waterloo, Waterloo, ON N2L 3G1, Canada \\
$^{2}$ Department of Physics and Astronomy, University of Waterloo, Waterloo, ON N2L 3G1, Canada \\
$^{3}$ Perimeter Institute for Theoretical Physics, 31 Caroline St. North, Waterloo, ON N2L 2Y5, Canada
}

\date{Accepted XXX. Received YYY; in original form ZZZ}

\pubyear{2021}

\begin{document}
\label{firstpage}
\pagerange{\pageref{firstpage}--\pageref{lastpage}}
\maketitle

\begin{abstract}
We present correction terms that allow delete-one Jackknife and Bootstrap methods to be used to recover unbiased estimates of the data covariance matrix of the two-point correlation function $\xi\left(\rr\right)$. We demonstrate the accuracy and precision of this new method using a large set of 1000 QUIJOTE simulations that each cover a comoving volume of $1\rm{\left[h^{-1}Gpc\right]^3}$. The corrected resampling techniques recover the correct amplitude and structure of the data covariance matrix as represented by its principal components to within $\sim10$\%, the level of error achievable with the size of the sample of simulations used for the test. Our corrections for the internal resampling methods are shown to be robust against the intrinsic clustering of the cosmological tracers both in real- and redshift space using two snapshots at $z=0$ and $z=1$ that mimic two samples with significantly different clustering. We also analyse two different slicing of the simulation volume into $\nsv=64$ or $125$ sub-samples and show that the main impact of different $\nsv$ is on the structure of the covariance matrix due to the limited number of independent internal realisations that can be made given a fixed $\nsv$.
\end{abstract}




\section{Introduction}

The two-point correlation function of the galaxy distribution contains a wealth of information about the gravity-driven growth of cosmological structure and the expansion rate of the Universe. As such the two-point correlation function and the power spectrum, its Fourier-space counterpart, have become an essential statistics to measure and fit with models of galaxy clustering expected in galaxy redshift surveys \citep[e.g.][]{Alam-BOSS-DR12,Alam-eBOSS-DR16}. 

Ongoing and planned experiments such as Euclid \citep{laureijs_euclid_2011} and the Dark Energy Spectroscopic Instrument (DESI; \citealt{desi_collaboration_desi_2016-1,desi_collaboration_desi_2016}) will provide an unprecedented amount of data enabling a percent level precision on the estimates of the key cosmological parameters. As with any experiment, the data analysis requires an equally accurate and precise estimate of the data covariance matrix, a key ingredient of model fitting. In cosmological surveys this is a challenging task as we are provided with only one realisation of the observed data.

Three different techniques have been proposed in the past literature to obtain estimates of the data covariance matrix: i) use of simulated datasets that provide an ensemble of synthetic replicas of the real observations that mimic the underlying probability distribution of the observed data \citep[e.g.][]{Zhao_2021}; ii) Analytical covariances that rely on a number of assumptions and theoretical models of the galaxy clustering in the Universe \citep[e.g.][]{Blake-cov}; iii) Internal estimators that resample directly the observed data using specific schemes to generate multiple copies of the observations \citep[e.g.][]{Norberg-Jackknife}. Each of these techniques has its pros and cons: For (i), accurate simulations of the non-linear effects are computationally expensive. Fast approximate simulations are usually employed to provide thousands of copies of real data. However they can fail to accurately predict the non-linear effects in the growth of structure, and their accuracy on small scales is debatable \citep[e.g.][]{Zhao_2021}. The recovered matrix is drawn from a Wishart distribution, and this contributes to the recovered model parameter errors \citep{hartlap07,DS13,Percival14,SH16,Percival21}. For the analytical calculations of the covariance matrix required for (ii), the simplifications required to be able to calculate the covariance may result in a biased estimate of the data covariance matrix, and the inclusion of window function and other effects is complicated \citep{Howlett2017,Blake-cov,Li2019}. Internal estimators on the other hand use appropriate resampling of the observed data and as such are less prone to unknown physics. However, they are found to provide estimates of the covariance matrix that can overestimate the true covariance matrix by anywhere between 25-60\% \citep{Norberg-Jackknife,Friedrich16,Favole2020}. Schemes to provide better convergence are generally complicated, such as the marked point bootstrap method of \citet{Loh2008}. In this paper we focus on (iii), exploring the application of the Jackknife and Bootstrap methods to analyses of the correlation function. 

Complications due to the window function make internal methods difficult to apply to the power spectrum, where subsampling would change the window effects (e.g. \citealt{wilson17}). This is not the case for the correlation function, where the estimator removes window effects, although one still have to be careful to subsample in a way that preserves volume scaling. Even so, the application is not straightforward because of the pairwise nature of the measurement \citep{Friedrich16}, which is the effect that we concentrate on here. The removal of subsets of the data to provide scale-able estimates of the variance is inherently a linear process, and assumes that the response to the removal of a fraction $1/\nsv$ of the data increases the variance by a factor $\nsv/(\nsv-1)$. For pair counts the situation is complicated by cross-pairs between regions being weighted in different ways. \citet{Friedrich16} investigated this in detail, considering different methods to correct for auto- and cross-pairs, but without finding a solution that works in all situations. The fundamental issue is how to weight for different numbers of auto- and cross-pairs in each bin in pair separation. We take up this problem, and show that a scaling of the pair counts leads to an unbiased estimator, accurate on large scales.

Throughout, we use the Landy-Szalay estimator \citep{LS-1993} to compute the two-point correlation function
\begin{equation}
    \xi(\rr) = \frac{\rm{DD}(\rr)-2\ DR(\rr)}{\rm{RR}(\rr)}+1, \label{eq:LS}
\end{equation}
where DD, DR and RR denote the data-data, data-random and random-random pairs, respectively. Given the regular geometry and periodic boundary conditions of the N-body simulations we could in principle use the natural estimator $(\DD/\RR-1)$ and analytically infer the $\RR$ term in order to estimate the two-point correlation function from the full-box simulations. This would also be computationally more efficient than using the Landy-Szalay estimator since we can use analytical expression to evaluate $\RR$ pair counts, typically the bottleneck in the measurement of the correlation functions. However we decided to use Eq.~\eqref{eq:LS} both for the measurements from the full-box simulations as well as from its Jackknife and Bootstrap realizations. This is due to the fact that the Jackknife and Bootstrap resamplings produce peculiar geometries of the sample for which Eq.~\eqref{eq:LS} provides the optimal estimator. In performing the pair counts in Eq.~\eqref{eq:LS} we also do not make use of the boundary conditions. This allows us to test the methods proposed in this paper in a similar setup to that used to estimate the two-point correlation function in galaxy redshift surveys.

We build a catalogue of un-clustered random points $\sim 25$ times the number of halos in the simulated samples. The data-data (DD), data-random (DR) and random-random (RR) pair-counts in Eq.~\eqref{eq:LS} are normalized to the corresponding total number of pairs in the sample. For estimates using the full-box simulations, i.e. without any resampling, with $N_h$ dark matter halos and $N_r$ random particles, the normalisation factors for $\DD$, $\RR$ and $\DR$ pair counts are $\rm N_{DD} = N_h(N_h -1)/2$, $\rm N_{RR} = N_r(N_r -1)/2$ and $\rm N_{DR} = N_h N_r$, respectively.

We consider the real-space correlation function $\xi(r)$, assumed to be a function of the pair separation only. We bin the pair-separations into $n_b=30$ linear bins of $5\rm h^{-1}Mpc$ between 0-150$\rm h^{-1}Mpc$. We also consider the multipoles of the redshift-space correlation function, calculated from the correlation function measured as a function of the redshift-space pair separation $s$ and the cosine $\mu$ of the angle between the pair-separation and the line-of-sight. We bin $s$ using the same scheme used for the real-space pair separation $r$ while $\mu$ is binned into $n_\mu=100$ linear bins between 0 and 1. Given the redshift-space two-point correlation function $\xi^{(s)}\left(s,\mu\right)$ its multipoles are obtained through,
    \begin{equation}
        \xi^{(\ell)}\left(s_i\right) = \left(2\ell+1\right)\sum_{j=1}^{n_\mu}\xi^{(s)}\left(s_i,\mu_j\right)L_\ell\left(\mu_j\right)\Delta\mu, \label{eq:mps}
    \end{equation}
where $L_\ell$ is the Legendre polynomial of order $\ell$ and $\Delta\mu=0.01$. In this work we limit our analysis to the first three even multipoles corresponding to $\ell=0,2,4$.

\section{QUIJOTE Simulations and covariance comparison}\label{sec:simulations}

\begin{figure}
    	\centering
		\includegraphics[width=\columnwidth]{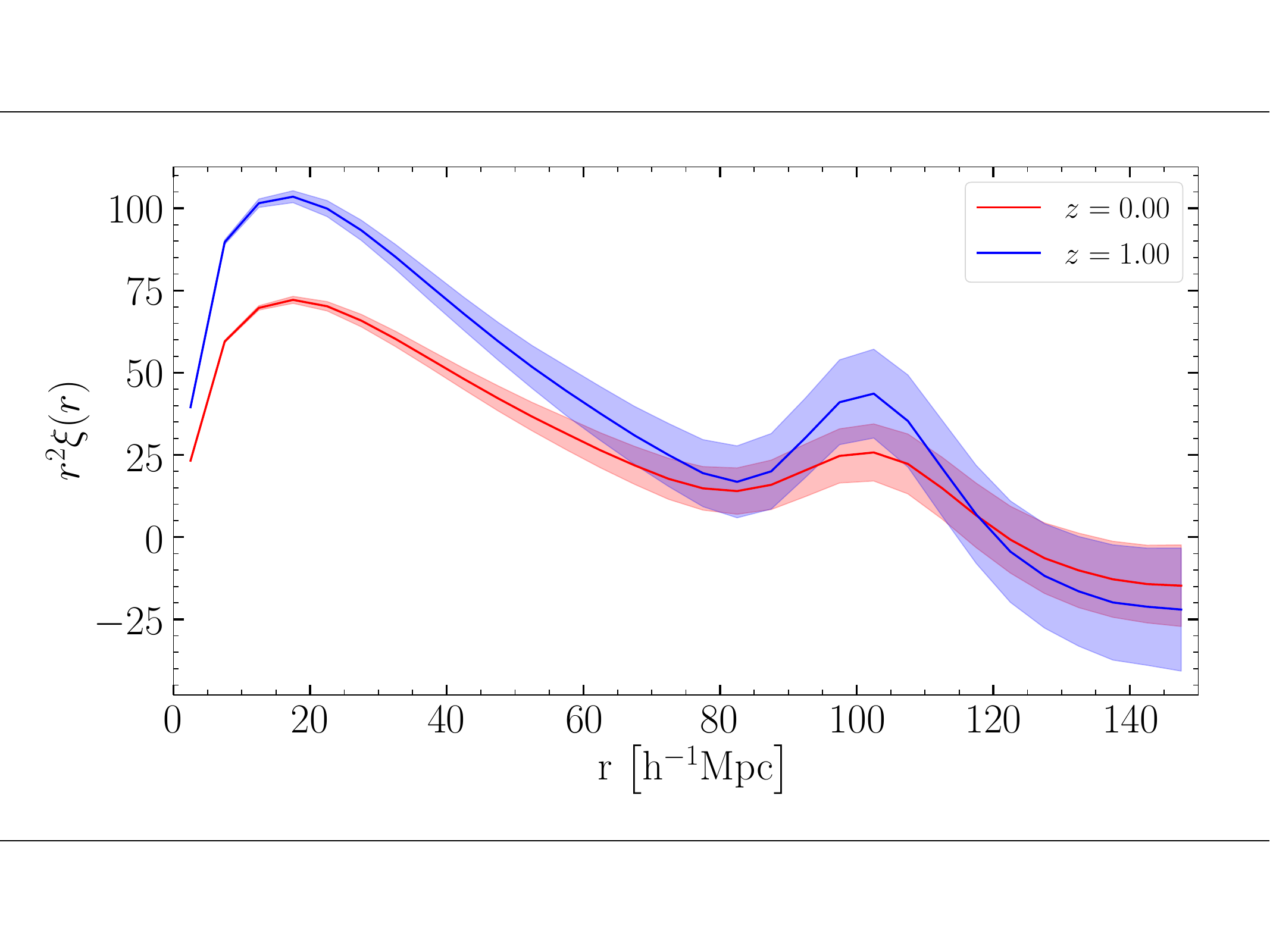}
		\caption{Real-space two-point correlation function measured from 1000 Quijote simulations.}\label{fig:xir_sims}
\end{figure}

We use a set of 1000 QUIJOTE dark matter N-body simulations to perform the analysis in this work. These simulations cover a comoving volume of $1\mhgpcc$ with periodic boundary conditions and use a flat $\Lambda$CDM cosmology consistent with Planck18 constraints. The cosmological parameters are set to $\left\{\Omega_m,\Omega_b,h,n_s,\sigma_8,w\right\}=\left\{0.3175, 0.049, 0.6711, 0.9624,0.834,-1\right\}$. The mass resolution of these simulations is $m_p=6.57\times10^{11}\mhmsun$. For our analysis we use the halo catalogues built using the Friends-of-Friends halo finder. The minimum number of particles in each halo is set to be 20 resulting in a halo mass $M_h>1.31\times10^{13}\mhmsun$. In particular we use two snapshots at z=0 and z=1 to test the dependence of our results on the intrinsic clustering of the sample. The halo catalogues contain on average $n_h\sim4\times10^5$ and $n_h\sim2\times10^5$ dark matter haloes at redshift $z=0$ and $z=1$, respectively. The two-point correlation function $\xi(r)$ in real-space averaged over the set of 1000 simulations at the two snapshot redshifts are shown in Fig.~\ref{fig:xir_sims}. 

Given the two-point correlation function measurements in $\nb$ separation bins obtained from $\ns$ independent datasets, the data covariance between the measurements in bin $i$ and bin $j$ can be estimated as,
\begin{equation}
    C_{ij}=\frac{1}{\ns-1}\sum_{k=1}^{\ns}\left[\xi^{[k]}_i - \langle\xi_i \rangle\right]\left[\xi^{[k]}_j - \langle\xi_j \rangle\right], \label{eq:covariance}
\end{equation}
where $\xi^{[k]}_i$ is the estimate in pair-separation bin $i$ made from $k$-th dataset while $\langle \xi_i\rangle$ denotes the corresponding ensemble average. This estimate is drawn from a Wishart distribution with $n_s-1$ degrees of freedom and scale matrix $\Sigma/(n_s-1)$, where $\Sigma$ is the true covariance. The variance for this distribution is 
\begin{equation}  
    {\rm Var}(C_{ij})
    =\frac{1}{\ns-1}\left(\Sigma_{ij}^2+\Sigma_{ii}\Sigma_{jj}\right)\,.
\end{equation}
In general, we will be comparing this estimate, which we refer to as the reference covariance matrix, against the covariance calculated using internal methods averaged over the same simulations. 

For a covariance matrix created from Jackknife realisations, this matrix can be considered to have been drawn from a Wishart distribution with $\nsv-1$ degrees of freedom \citep{Efron1980,ShaoWu1989}, provided that the subsamples are independent. In practice, this is not the case for the correlation function thanks to cross-pairs, and the effective degrees of freedom for the distribution from which the covariance matrix is drawn will be smaller than $\nsv$. We consider Jackknife or Bootstrap realisations with $\nsv=64$, or $\nsv=125$, while we have $\ns=1000$ simulations, and so the error associated with the reference covariance is small compared with that from an individual Jackknife realisation. However, we average results calculated using internal methods across all simulations, reducing the error to a level below that on the reference covariance. Unfortunately, the averaged result will also be correlated with the reference covariance, as they were calculated from the same set of mocks. Furthermore, we cannot easily estimate the error on the difference given that we only have one realisation of the reference covariance. 

In most plots we therefore plot errors appropriate to a single Jackknife realisation in order to demonstrate any offsets relative to this error, which is the most important for any individual analysis and allows us a consistent approach for all plots. This is larger than the error on the reference cosmology, and so we are justified in ignoring this when plotting these errors. As we do not know exactly the degrees of freedom for the Wishart matrix from which the Jackknife distribution should be drawn, we estimate errors on the Jackknife covariance matrix from the scatter across mocks.

In order to make the comparison between two estimates of the covariance matrix easier, we first re-write the covariance matrix $C$ in terms of its diagonal components $\bm\sigma$ and the correlation matrix $R$,
\begin{equation}
    C = {\bm\sigma}R{\bm\sigma},
\end{equation}
with $\sigma_{ij}^2=C_{ij}\delta^K(i-j)$ and $R_{ij}=C_{ij}/\left(\sigma_{ii}\sigma_{jj}\right)$. The term ${\bm\sigma}$ is the variance of the measurements in a given bin while the correlation matrix $R$ quantifies the correlation between different measurement bins and captures the structure of the covariance matrix. We further decompose the correlation matrix $R$ into its principal components,
\begin{equation}
    R{\bm v}_i = \lambda_i{\bm v}_i \label{eq:pca}
\end{equation}
with $\lambda_i$ and ${\bm v}_i$ being the i-th eigenvalue and eigenvector of the correlation matrix $R$. A comparison between two estimates of the covariance matrix now translates into an element-wise comparison of its diagonal elements, namely the variances ${\bm\sigma}$, eigenvalues $\lambda$ and eigenvectors ${\bm v}$. In the rest of the paper we will use the normalised eigenvalues such that their sum is equal to 1 but keep the notation for the normalised eigenvalues the same as in Eq.~\eqref{eq:pca}.

\section{Internal covariance estimation methods}\label{sec:methods}

We will consider two internal estimators of the data covariance matrix - Bootstrap and delete-one Jackknife (or simply Jackknife). Both rely on making multiple copies of the observed dataset using different resampling methods. To do this, data are first split into $\nsv$ non-overlapping sub-samples that are then resampled according to the Bootstrap or Jackknife methods. In this work we slice the full-box simulations into identical cubic sub-samples. To test the impact of the number of sub-volumes $\nsv$ on the estimated covariance matrix we use two values of $\nsv=64$ and $\nsv=125$. A smaller number of sub-volumes would result in a much smaller number of Jackknife realisations that can be drawn from each simulation providing noisy or even singular estimates of the data covariance matrix depending on the number of measurement bins $\nb$. A larger number of sub-samples requires an amount of computational resources and time much larger than that available for this work.

\subsection{Jackknife}

The delete-one Jackknife realisations are built removing only one of the $\nsv$ sub-samples at a time, and calculating the correlation function for the remaining data. Thus we have $\nsv$ Jackknife realisations and each uses a smaller fraction of the total volume, specifically equal to $(\nsv-1)/\nsv$. To account for this the Jackknife estimate of the covariance matrix is
    \begin{equation}
        C_{ij}^{\rm \left(jk\right)}=\frac{\nsv-1}{\nsv}
        \sum_{k=1}^{\nsv}\left[\xi^{[k]}_{i,\rm jk} - \bar{\xi}_{i,\rm jk}\right]\left[\xi^{[k]}_{j,\rm jk} - \bar{\xi}_{j,\rm jk}\right]. \label{eq:jk}
    \end{equation}
In Eq.~\eqref{eq:jk},
    \begin{equation*}
        \bar{\xi}_{i,\rm jk}=\frac{1}{\nsv}\sum_{k=1}^{\nsv}\xi^{[k]}_{i,\rm jk},
    \end{equation*}
is the mean estimate from $\nsv$ Jackknife realisations. The pre-factor $\left(\nsv-1\right)/\nsv$ in Eq.~\eqref{eq:jk} correctly accounts for the reduction in the sample size only for the one-point statistics. For the two-point correlation function, removing one of $\nsv$ sub-samples does not guarantee, in general, a Jackknife estimate as drawn from an effective sample $(\nsv-1)/\nsv$ in size compared to the full sample (see section~\ref{sec:weighting}). 

In analogy to the Bootstrap resampling we can view the delete-one Jackknife resampling as a weighting of the $\nsv$ sub-samples where for the $i$-th Jackknife realisation all sub-volumes are weighted 1 except the sub-sample indexed $i$ that is weighted 0. In particular, for the $i$-th Jackknife realisation the vector of weights $\mathbf{w}^{\left(i,\rm jk\right)}$ is $w_{j=1,\dots,\nsv}^{\left(i,\rm jk\right)}=\left[1-\delta^D(i-j)\right]$. The $\nsv$ Jackknife realisations built in this way form a set of linearly independent realisations given the linear independence of the weight vectors $\mathbf{w}^{\left(i,\rm jk\right)}$. As a result the covariance matrix is guaranteed to be non-singular as long as the number of bins $\nb$ is smaller than the number of Jackknife realisations $\nsv$.

\subsection{Bootstrap}

Bootstrap realisations are drawn by randomly resampling $\nsv$ sub-samples with replacement, i.e. each sub-sample can be drawn more than once in the same realisation. Simply put, to generate a Bootstrap realisation of the data we assign a random integer weight to each sub-sample. Naturally, all objects with a given sub-sample also inherit its weight. Each realisation is thus characterised by a vector of weights $\mathbf{w}$ of size $\nsv$. However, for a given realisation, the elements of the vector $\mathbf{w}$ cannot take arbitrarily large values. In this work we set the constraint,
    \begin{equation}
        \sum_{i=1}^{\nsv} w_i=\nsv, \label{eq:bs_constraint}
    \end{equation}
i.e. the sum of individual weights $w_i$ must sum to the total number of sub-samples $\nsv$.

Given $\nbs$ Bootstrap realizations of the data, the related covariance matrix is estimated as
    \begin{equation}
        C_{ij}^{\left(\rm bs \right)}=\frac{1}{\nbs-1}\sum_{k=1}^{\nbs}\left[\xi^{[k]}_{i,\rm bs} - \bar{\xi}_{i,\rm bs} \right]\left[\xi^{[k]}_{j,\rm bs} - \bar{\xi}_{j,\rm bs}\right], \label{eq:bs}
    \end{equation}
where $\xi^{[k]}_{i,\rm bs}$ is the $k$-th Bootstrap estimate in bin $i$ and,
    \begin{equation*}
        \bar{\xi}_{i,\rm bs}=\frac{1}{\nbs}\sum_{k=1}^{\nbs}\xi^{[k]}_{i,\rm bs}
    \end{equation*}
is the mean estimate from $\nbs$ Bootstrap realizations. Unlike the Jackknife method, we can have $\nbs\ne\nsv$.

\section{Weighting Schemes for internal covariance estimates}\label{sec:weighting}

As described in Section~\ref{sec:methods}, internal resampling methods can be seen as a slicing of the original data and weighting the resulting $\nsv$ sub-samples following specific prescriptions. Here we address an ambiguity in method because we are dealing with pairwise weights, as required to perform internal estimates of the two-point correlation function. This complicates the analysis, compared with, for example individual weights $w_i$ of the $\nsv$ sub-samples. Given a set of weights $w_{i=1,\dots,\nsv}$ for individual sub-samples, there is no unique choice in assigning weights to pairs of objects. This holds for auto-pairs where both objects forming a pair reside in the same sub-sample, and for cross-pairs where the two objects lie in two distinct sub-samples. As we will see in Section~\ref{sec:results}, the way weights are assigned to pairs of objects has a major impact on the accuracy of internal estimates of the data covariance matrix. We now consider three different ways of assigning weights to pairs of objects given the individual weights $w_i$ and $w_j$ of the sub-samples they reside in. We then use a simple model for the covariance as a sum of covariances from auto- and cross-pairs, to show how the simple Jackknife results can be scaled to allow for the differences. This then leads us to consider a weighting scheme designed to match the size of effect on auto- and cross-pairs. We only consider weighting schemes that remove all auto-pairs from within a subvolume, and weight or remove cross-pairs for which one galaxy lies in the same subvolume. These weights are applied to $\DD$, $\DR$ and $\RR$ pairs, so for each set of weights, the match between expected pair numbers is maintained.

\subsection{$\vmult$ Weighting} \label{sec:v0}

This is perhaps the most widely used method to assign weights to pairs of objects where weights of the two objects forming the pair are multiplied together to obtain their pairwise weight. In this weighting scheme, if two objects reside in sub-samples $i$ and $j$ which individual weights are $w_i$ and $w_j$, respectively, the corresponding pair will be weighted by $w_{ij}$ where,
\begin{equation}
    w_{ij}^{\vmult} = w_i w_j\,. \label{eq:v0}
\end{equation}
This weighting scheme fails to match the contributions to the covariance matrix from cross-pairs. In the Jackknife realisations both the number of auto pairs and the volume contributing to their counts are reduced by $1/\nsv$ with respect to the full-data estimates. Since this reduction is accounted for by the prefactor in Eq.~\eqref{eq:jk} the Jackknife resamplings provide a correct estimate of the shot-noise and cosmic-variance contribution from the auto pairs. For the cross-pairs on the other hand, the number and volume are reduced on average by a factor of $2/\nsv$ overestimating the shot-noise and cosmic variance contributions to the covariance matrix from the cross pairs. For the Bootstrap resamplings, these same considerations are difficult to generalise given the different ways the weights can be assigned to $\nsv$ sub-samples. In general, Bootstrap estimates of the covariance matrix made using $\vmult$ weights will be strongly biased. To see this consider a toy example where one sub-sample is weighted 0 and the constraint $\sum_{i=1}^{\nsv}w_i=\nsv$ requires another sub-sample to be weighted 2. On average the net effect is to increase the expected auto pairs by scaling by a factor of $1+2/\nsv$ and reduce the number of cross-pairs by scaling by a factor $1-2/\left(\nsv\left(\nsv-1\right)\right)$. Given these complications we have considered other ways of weighting pairs. A simple rescaling of the Jackknife error estimates using $\vmult$ weighting, that accounts for the mismatch in the cross-pairs, is discussed in Section~\ref{sec:rescaling}.

\subsection{$\vmean$ Weighting} \label{sec:v1}
Changing the weighting scheme can change the relative contributions of auto- and cross-pairs. To see this, rather than using $\vmult$, consider weighting instead by a function,
\begin{equation}
    w_{ij}^{\vmean} = \frac{w_i + w_j}{2}\,. \label{eq:v1}
\end{equation}
This changes the balance between auto- and cross-pairs compared with Eq.~\ref{eq:v0}. For Jackknife realisation $i$, instead of removing cross pairs with one galaxy in subregion $i$, we instead weight them by $1/2$, decreasing the importance of cross-pairs in the realisations compared to auto pairs, and increasing their contribution to the variance. We discuss this further in Section~\ref{sec:rescaling}.

One interesting property of the $\vmean$ weighting is that for the Bootstrap resampling it limits the samples to be a linear combinations of the Jackknife realisations. As defined in Eq.~\eqref{eq:v1}, any vector of pairwise weights $w_{ij}^{\vmean}$ is a linear combination of the weight vectors $w_i$ and $w_j$. Since there are $\nsv$ sub-samples in total, the maximum number of independent vectors that form a basis for the vectors $w_i$ and $w_j$ is equal to $\nsv$. One such basis is given by the weight vectors associated with the Jackknife realisations, and so it is easy to see that with the $\vmean$ weighting any Bootstrap realisation (or equivalently its corresponding weights $w_{ij}$) can be expressed as a linear combination of the Jackknife realisations. For this weighting scheme, we therefore expect similar results for Bootstrap and Jackknife methods.

\subsection{$\vgeom$ Weighting} \label{sec:v2}

We also test a weighting scheme that uses the geometric mean of the individual sub-samples weights to compute pairwise weights,
\begin{equation}
    w_{ij}^{\vgeom} = \left[w_i w_j\right]^{1/2}.   \label{eq:v2}
\end{equation}
Both $\vmean$ and $\vgeom$ provide the same number of auto-pairs ($i=j$) and the corresponding volume. However, they differ in the case of cross-pairs. As for the  $\vmult$ weighting, with the $\vgeom$ weighting we systematically remove all pairs if at least one object lies in one of the removed sub-samples which weight is set to 0. Finally we note that for the Jackknife resampling where the weight $w_i$ of each sub-sample is either 0 or 1, $\vmult$ and $\vgeom$ weightings provide identical results. For the Jackknife resampling we will thus present results only from $\vmult$ and $\vmean$ weighting schemes.

\subsection{The effect of differences in weighting auto- and cross-pairs on Jackknife results}  \label{sec:rescaling}

As discussed above, the different weighting schemes weight auto- and cross-pairs by different amounts, and this will lead to different final variance estimates. In this section, we analyse this in more detail and outline a correction for it. First, let us consider the Jackknife method concentrating on pair counts, which drive the error on the correlation function. 

Given the slicing of the original data into $\nsv$ sub-samples, we denote by $\DD_{a,k}$ and $\DD_{c,k}$ the auto- and cross-pairs respectively removed in Jackknife realisation $k$ for a given bin of $\xi$. $\overline{\DD}_a$ is the average of $\DD_{a,k}$ over all of the Jackknife realisations and $\overline{\DD}_c$ the average of $\DD_{c,k}$. We can then write the total number of auto- and cross-pairs in the full sample from which we have made the Jackknife realisations
\begin{eqnarray}
    DD^{\rm tot}_a &=& \nsv\overline{DD}_a\,, \label{eq:dd_auto}\\
    DD^{\rm tot}_c &=& \frac{\nsv}{2}\overline{DD}_c\,. \label{eq:dd_cross}
\end{eqnarray}
Here we have used the fact that all of the weighting schemes considered removed all auto-pairs within a subvolume, and weighted or removed all cross-pairs where one galaxy lies within the volume. Because the number of cross-pairs affected is approximately twice that of auto pairs, the scaling in Eq.~\ref{eq:dd_cross} is different for the two types. 

We now consider the effect of the Jackknife method on the relative numbers of auto- and cross-pairs. This ignores the correlation between their numbers. Because we focus on the numbers of pairs we also expect that this analysis works better for the shot-noise contribution, rather than the sample variance contribution to the covariance matrix.

Each Jackknife realisation, for any of the weighting schemes considered above, therefore contains 
\begin{equation} \label{eq:theta_ai}
    \theta_{a,k}=\frac{1}{\nsv-1}\left(\nsv\overline{DD}_a-DD_{a,k}\right)\,,
\end{equation}
normalised auto-pairs in the pair-separation bin of interest. We now define $\overline{\theta}_a$ as the average number of pairs over all Jackknife samples associated with a particular survey, and consider variation of the Jackknife estimates around this. So we have that $\theta_{a,k}-\overline{\theta}_a=(\overline{DD}_a-DD_{a,k})/(\nsv-1)$. Squaring this and summing we have the standard Jackknife result for the variance on the number of auto-pairs 
\begin{equation}
    (\nsv-1)\sum_{k=1}^{\nsv}(\theta_{a,k}-\overline{\theta}_a)^2
        =\frac{1}{\nsv-1}\sum_{k=1}^{\nsv}(DD_{a,k}-\overline{DD}_a)^2\,.
\end{equation}
Noting that the final size of the sample is $\nsv$ times larger than removed in any Jackknife sample, and that the variance scales as $1/\nsv$, the variance estimate for the full sample if we only considered auto-pairs would be 
\begin{equation}  \label{eq:S_a}
    S_a = \frac{\nsv-1}{\nsv}
    \sum_{k=1}^{\nsv}(\theta_{a,k}-\overline{\theta}_a)^2\,,
\end{equation}
matching the form of Eq.~\ref{eq:jk}. In contrast, for the cross-pairs, for $\vmean$ and $\vgeom$ weighting, we have that 
\begin{eqnarray}
    \theta_{c,\vmean,k}&=&\frac{2}{\nsv-1}
        \left(\frac{\nsv}{2}\overline{DD}_{c,\vmean}-\frac{1}{2}DD_{c,\vmean,k}\right)\,,\\
    \theta_{c,\vgeom,k}&=&\frac{2}{\nsv-2}
        \left(\frac{\nsv}{2}\overline{DD}_{c,\vgeom}-DD_{c,\vgeom,k}\right)\,.
\end{eqnarray}
where $\theta_{c, k}$ is the cross-pairs equivalent of the $\theta_{a, k}$ in Eq.~\eqref{eq:theta_ai} defined for the auto pairs. The sub-script $\vmean$ ($\vgeom$) in $\theta_{c, \vmean, k}$ ($\theta_{c, \vgeom, k}$) refers to the weighting scheme. These are clearly different from Eq.~\ref{eq:theta_ai}, stemming from the different proportion of cross-pairs removed or weighted for each Jackknife realisation compared with auto-pairs. Working through the same derivation considered above for auto-pairs, and noting that for cross-pairs the final size of the sample is $\nsv/2$ times larger than removed in any Jackknife sample, we find that
\begin{eqnarray}
    S_{c,\vmean} &=&
    \frac{2(\nsv-1)}{\nsv}\sum_{k=1}^{\nsv}(\theta_{c,\vmean,k}
        -\overline{\theta}_{c,\vmean})^2\,,\label{eq:Svmean}\\
    S_{c,\vgeom} &=&
    \frac{(\nsv-2)^2}{2\nsv(\nsv-1)}\sum_{k=1}^{\nsv}(\theta_{c,\vgeom,k}
        -\overline{\theta}_{c,\vgeom})^2\,.\label{eq:Svgeom}
\end{eqnarray}
Note that the recovered values from these expressions are unaffected by correlations between $\theta_{c,k}$, which do not affect the variance recovered. These represent the scaling of the Jackknife results that we should have in order to obtain the correct covariances if we only considered cross-pairs. 

In practice, the Jackknife realisations are analysed including all pairs, and are scaled according to Eq.~\ref{eq:S_a}, such that the auto-pair component is correct and the cross-pair component is incorrect. We now consider a scaling of the result that will correct for this inconsistency. 

We make the approximation that the final covariance depends on the combination of variances from auto- and cross-pairs, assuming optimal combination. Defining $f$ as the fraction of pairs that are auto-pairs, to correct for the difference responses of Jackknife errors to auto- and cross-pairs, we need to rescale the Jackknife covariance calculated using the $\vmean$ weights by
\begin{equation}    
    \beta_{\vmean}(r) = f(r)+2\left[1-f(r)\right]\,, \label{eq:recsaling1}
\end{equation}
and the Jackknife covariance calculated using the $\vgeom$ or $\vmult$ weights by
\begin{equation}    
    \gamma_{\vgeom}(r) = f(r)+\frac{(\nsv-2)^2}{2(\nsv-1)^2}\left[1-f(r)\right]\,. \label{eq:recsaling2}
\end{equation}
As we will see in the next section, these approximate rescalings work well and are able to correct the Jackknife covariance estimates within the errors of our test data. For the Bootstrap estimates, given the discussion of $\vmean$ weighting in Section~\ref{sec:v1} we should expect that this rescaling also works well for Bootstrap realisations undertaken with this weighting scheme, and we find that this is empirically confirmed in Appendix~\ref{app:appendix_vmean}. For the other weightings this simple corrections is less successful for Bootstrap-based error estimates: this is not unexpected given that it has not been derived for the Bootstrap method.

\subsection{$\vmatch$ Weighting} \label{sec:v2}

We now consider, instead of correcting the variance after estimation, developing a weighting scheme that matches the size of effect on auto- and cross-pairs. Suppose that the effect on the auto-pairs is as considered in Section~\ref{sec:rescaling}, and consider the cross-pairs with a more general weighting scheme where we weight each pair by a factor $\alpha$. We would then have
\begin{equation}
    \theta_{c,\vmatch,k}=\frac{2}{\nsv-2\alpha}
        \left(\frac{\nsv}{2}\overline{DD}_{c,\vmatch}-\alpha DD_{c,\vmatch,k}\right)\,.
\end{equation}
The variance of this component is
\begin{equation}  
    S_{c,\vmatch} =
    \frac{(\nsv-2\alpha)^2}{2\alpha^2\nsv(\nsv-1)}\sum_{k=1}^{\nsv}(\theta_{c,\vmatch,k}
        -\overline{\theta}_{c,\vmatch})^2\,, \label{eq:Svmatch}
\end{equation}
which reduces to Eqns.~\ref{eq:Svmean} \&~\ref{eq:Svgeom} for $\alpha=1/2$ \& $\alpha=1$ respectively. In order to match the required scaling of the variance between auto- and cross-pairs we need to match this to Eq.~\ref{eq:S_a}, which requires
\begin{equation}
    \frac{(\nsv-2\alpha)^2}{2\alpha^2\nsv(\nsv-1)}=\frac{\nsv-1}{\nsv}\,.
\end{equation}
The important solution (for $\nsv\ge3$) is that
\begin{equation}
    \alpha=\frac{\nsv}{2+\sqrt{2}(\nsv-1)}\,.
\end{equation}
To summarise, for every Jackknife realisation $k$ we should remove all auto-pairs within subvolume $k$, and weight cross-pairs that have one galaxy in $k$ and one outside by $\alpha$. This will then scale the final variance estimate in the same way for both auto- and cross-pairs.

\section{Results} \label{sec:results}

In this section we compare internal estimates of the covariance matrix against the reference measured using Eq.~\ref{eq:covariance}. We start in Section~\ref{sec:results_baseline} by presenting the outcomes of our preferred, $\vmatch$ weighting scheme for the Jackknife resampling method (see Section~\ref{sec:rescaling}), slicing the $z=0$ data into $\nsv=125$ sub-samples. We call this our {\it baseline} method. In Section~\ref{sec:results_weightings} we compare our rescaling schemes using the Jackknife resampling method and show the effectiveness of the rescaling factors in Eqns.~\eqref{eq:recsaling1} and ~\eqref{eq:recsaling2} for correcting the Jackknife covariance matrix obtained using the $\vmean$ and $\vmult$ weightings. Section~\ref{sec:bootstrap} compares results from Jackknife and Bootstrap methods. The impact of the number of sub-samples $\nsv$ is investigated in Section~\ref{sec:slicing}. In Section~\ref{sec:clustering} we analyse the robustness of our method against the intrinsic clustering of the sample by comparing against catalogues at $z=1$.

\subsection{Analysis using $\vmatch$}  \label{sec:results_baseline}

\begin{figure}
    	\centering
		\includegraphics[width=\columnwidth]{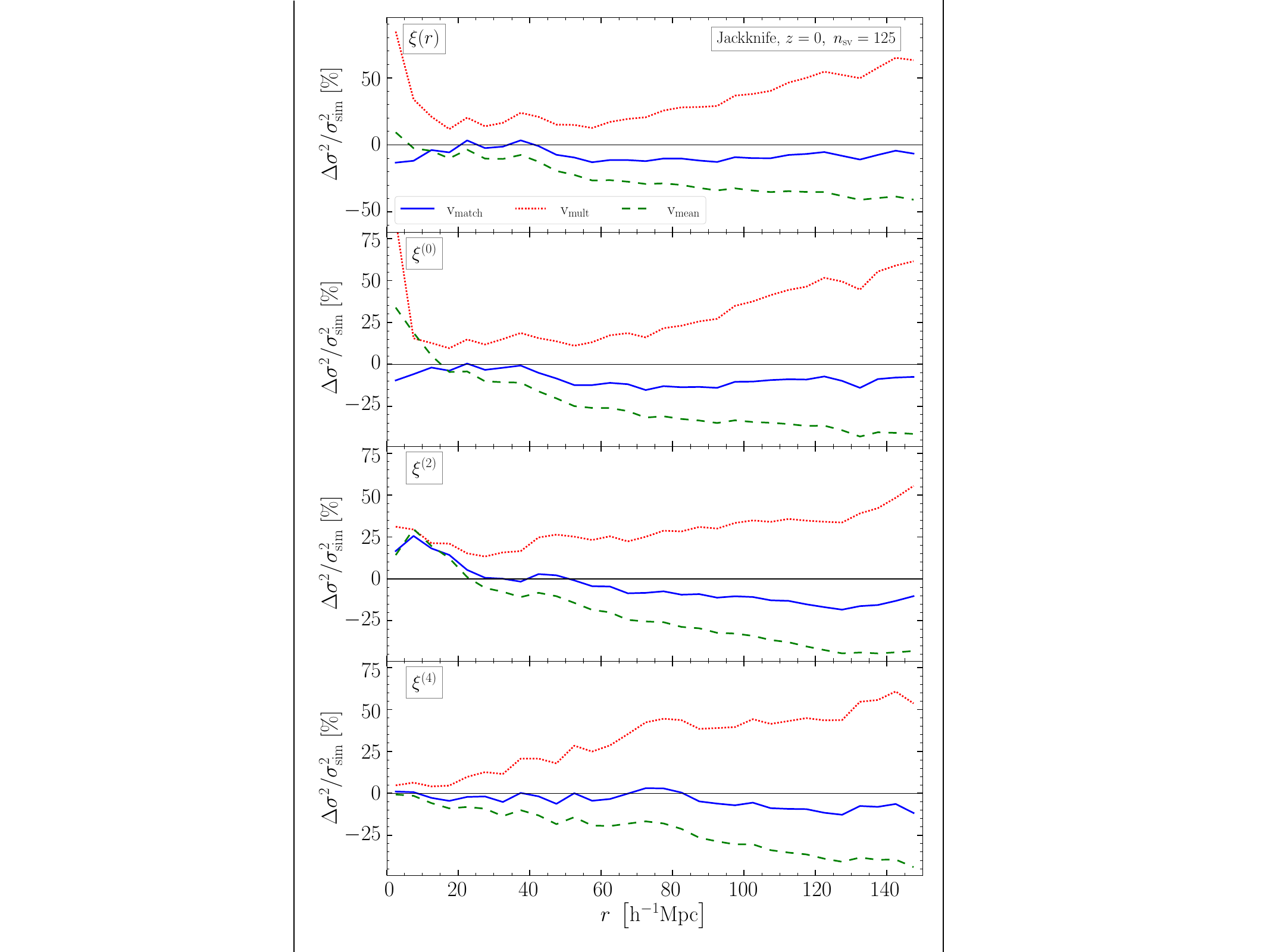}
		\caption{Relative difference between the Jackknife and reference estimates of the variances for the two-point correlation function. Jackknife estimates slice the data into $\nsv=125$ sub-samples. We show the results for the three weighting schemes: the standard $\vmult$ weighting (red dotted lines); the shot-noise correct $\vmean$ weighting (green dashed lines) and the optimal $\vmatch$ weighting (blue thick lines). Reference estimates are obtained using a set of $\ns=1000$ independent QUIJOTE simulations. Top panel: real-space correlation function. Bottom three panels: multipole moments of the redshift-space correlation function.}\label{fig:vars_xir_baseline}
\end{figure}

\begin{figure}
    	\centering
		\includegraphics[width=\columnwidth]{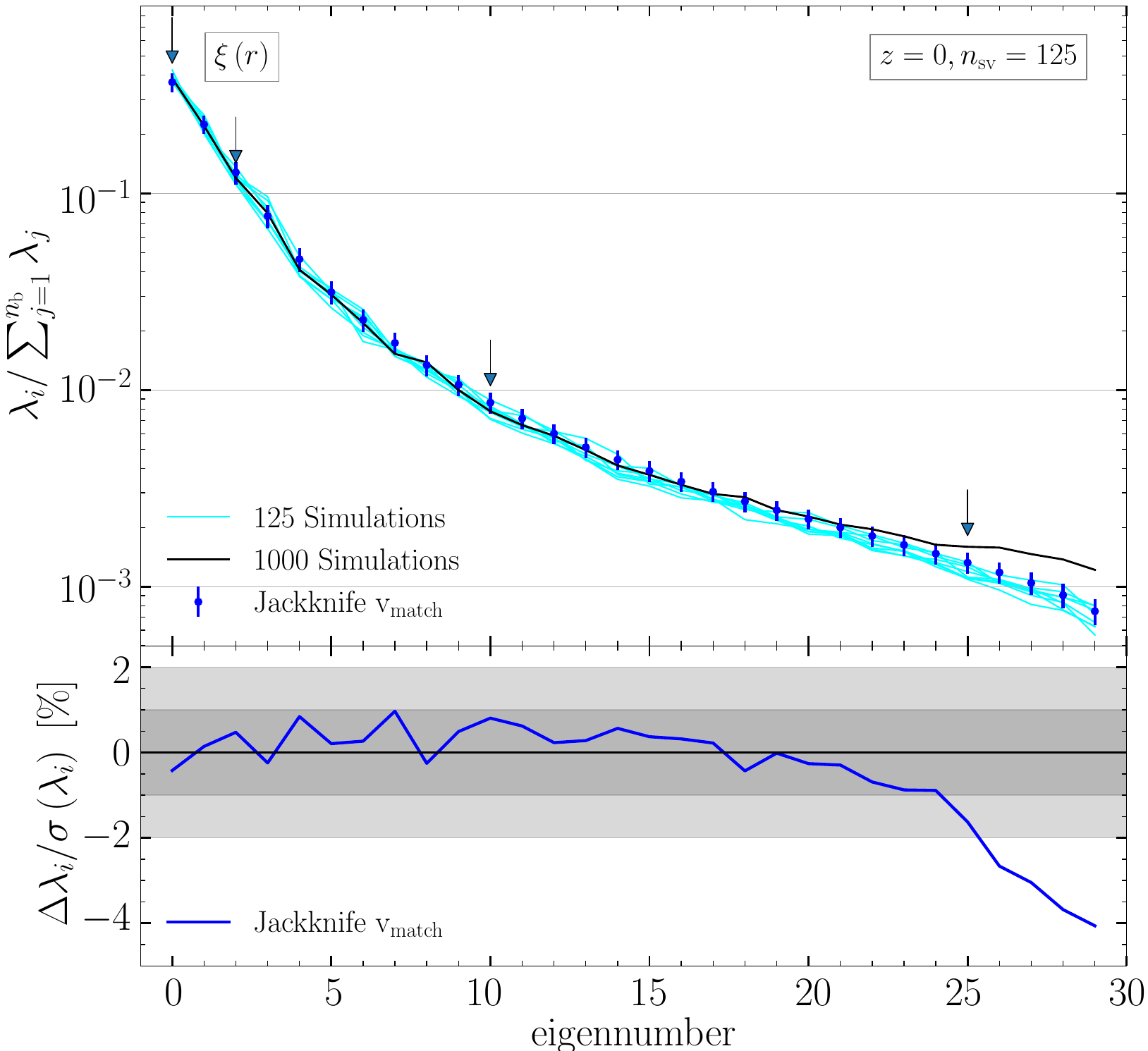}
		\caption{Top panel: internal (points with errorbars) and reference estimates from $\ns=1000$ QUIJOTE simulations (black thick line) of the normalised eigenvalues of the covariance matrix for the real-space two-point correlation function. For comparison we also show estimates using subsets of only $\ns=\nsv=125$ simulations each (cyan lines). Bottom panel: difference between internal and reference estimates in units of $1\sigma$ error on internal estimates for a single simulation. Horizontal shaded bands delimit the $1-$ and $2\sigma$ intervals. The arrows in the top panel highlight the components for which we show the corresponding eigenvectors in Fig.~\ref{fig:eigenvectors_xir_baseline}.} \label{fig:eigenvalues_xir_baseline}
\end{figure}

\begin{figure}
		\includegraphics[width=\columnwidth]{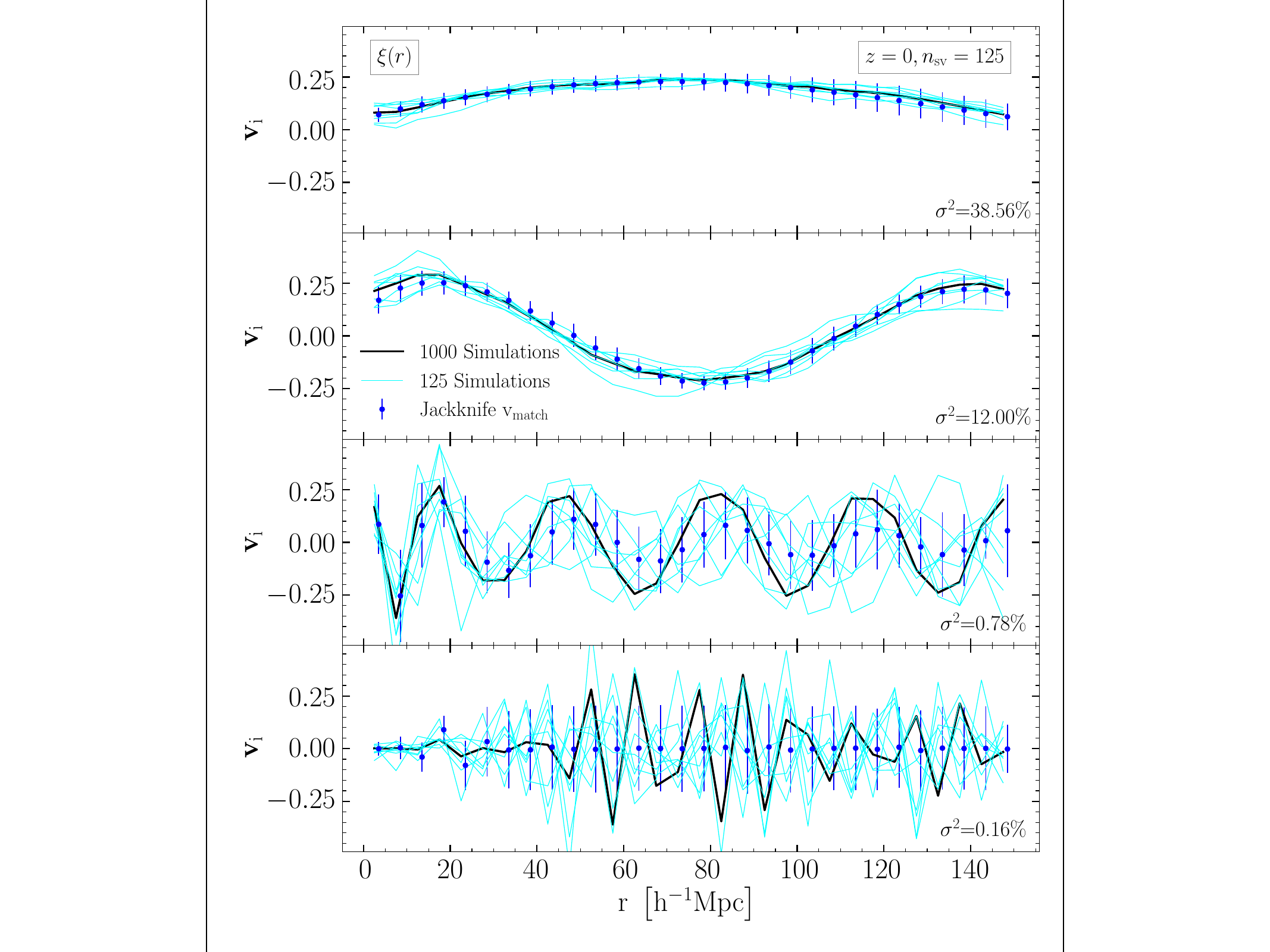}
		\caption{Baseline internal (points with errorbars) and reference estimates from $\ns=1000$ QUIJOTE simulations (black lines) of four different eigenvectors of the covariance matrix for the real-space two-point correlation function. Eigenvectors plotted here correspond to the components marked with vertical arrows in Fig.~\ref{fig:eigenvalues_xir_baseline}. For comparison we also show estimates using a number of simulations equal to the number of sub-samples $\nsv$ used to slice the data (cyan lines).}\label{fig:eigenvectors_xir_baseline}
\end{figure}

\begin{figure}
    	\centering
		\includegraphics[width=\columnwidth]{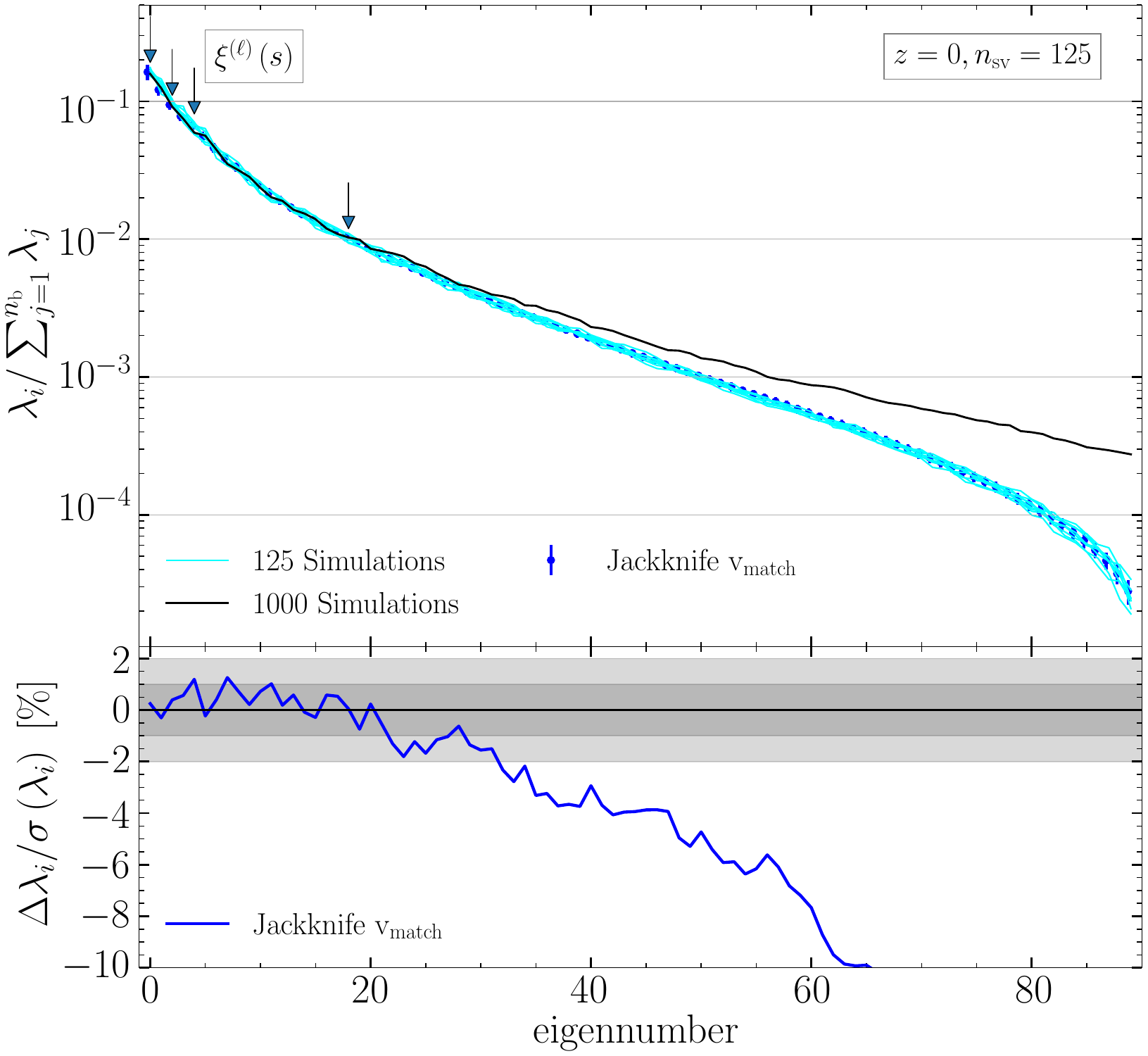}
		\caption{Same as in Fig.~\ref{fig:eigenvalues_xir_baseline} but here for the baseline analysis for the multipole moments of the redshift-space two-point correlation function. Note the much larger number of eigenvalues compared to the real-space case. Each multipole measurement uses the same linear binning for the pair separation as the one used in real space, hence the dimension of the data vector in redshift space is three times larger than in real space. Although the discrepancy between internal and reference estimates can reach values as low as $-50$ time the standard deviation we have truncated the lower limit in the y-axis of the bottom panel to $-10$ for clarity. The arrows highlight the components for which we show the corresponding eigenvectors in Fig.~\ref{fig:eigenvectors_mps_baseline}.}\label{fig:eigenvalues_mps_baseline}
\end{figure}

\begin{figure}
    	\centering
		\includegraphics[width=\columnwidth]{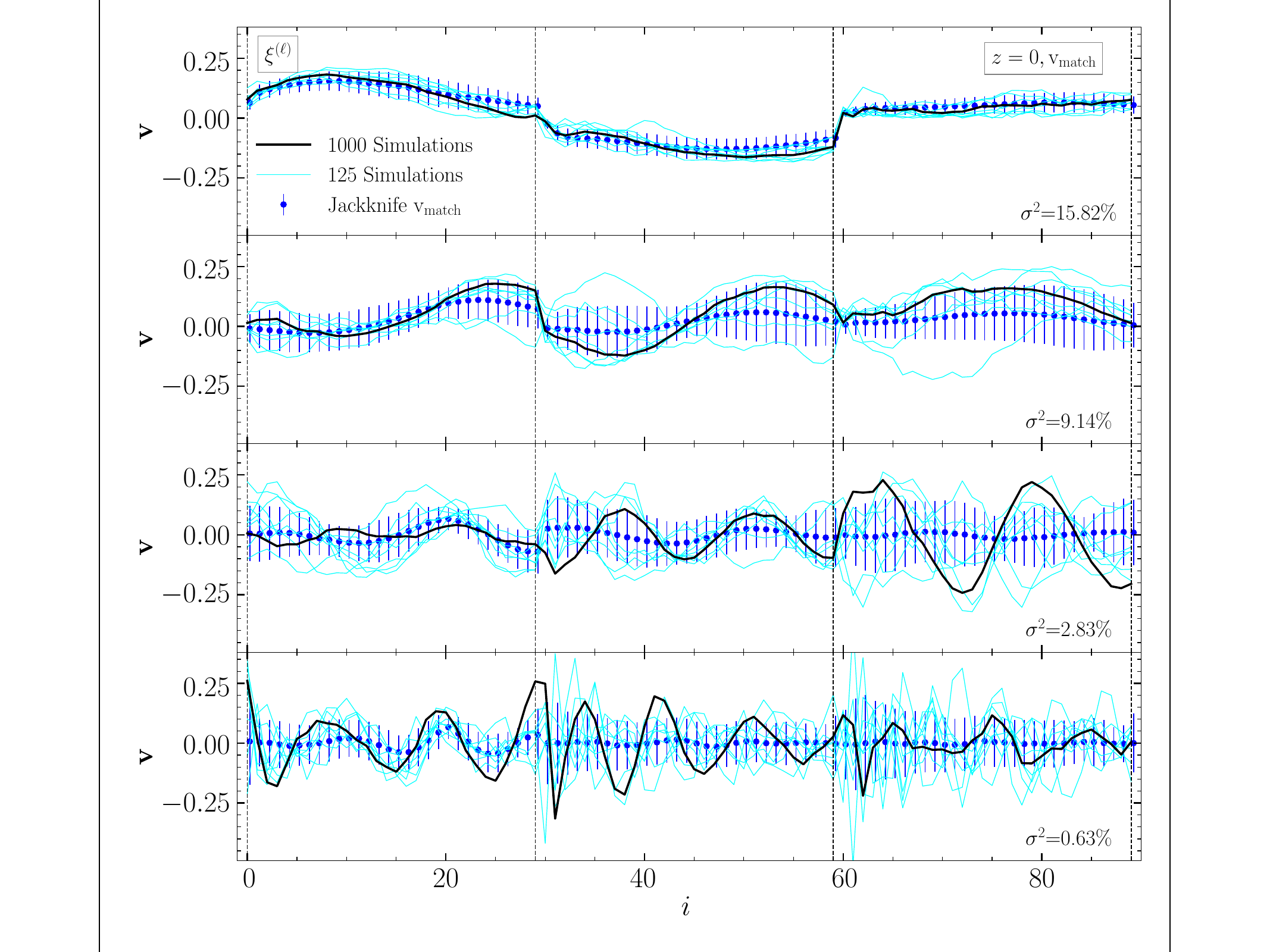}
		\caption{Same as in Fig.~\ref{fig:eigenvectors_xir_baseline} but here for the baseline analysis for the multipole moments of the redshift-space two-point correlation function. Different panels show eigenvectors for the components highlighted with vertical arrows in Fig.~\ref{fig:eigenvalues_mps_baseline}. Vertical dashed line delimit the parts of the full eigenvector related to the three multipoles.}\label{fig:eigenvectors_mps_baseline}
\end{figure}

We start analysing the results of the baseline configuration: Fig.~\ref{fig:vars_xir_baseline} shows the accuracy of the Jackknife estimates of the variances obtained using the $\vmatch$ weighting scheme for the real-space correlation function and the redshift-space multipoles. Overall, the $\vmatch$ weighting allows us to successfully recover the reference variances over the full range of scales explored here. The raw $\vmean$ weighting underestimates while the raw $\vmult$ overestimates the reference variances at large scales, matching a similar analysis by \citet{Friedrich16}. The magnitude of the offset in the case of raw $\vmean$ and $\vmult$ weightings can reach values up to $\sim50\%$. 

Figure~\ref{fig:eigenvalues_xir_baseline} shows the comparison for the normalised eigenvalues for the real-space correlation function, that represent the relative contribution to the total variance from the corresponding principle component, for our baseline analysis in real space. We find an excellent agreement between the internal and reference estimates for the main components up to the eigennumber of $24$ (25 out of 30 components) that jointly account for $\sim99\%$ of the total variance. This fraction is calculated as the cumulative contribution from all components up to a given eigennumber $\bar{n}$ as $\sum_{i=1}^{\bar{n}}\lambda_i$ with $\lambda_i$ being the $i$-th normalised eigenvalue. For less important components, internal methods underestimate the reference values by more than $2\sigma$ due to a lower statistical power of the resamplings compared to that of 1000 independent simulations. There is very good agreement between internal estimates, that resample the data using $\nsv=125$ sub-samples, and those obtained using subsets of $\ns=125$ independent simulations each (cyan lines in Fig.~\ref{fig:eigenvalues_xir_baseline}). This indicates that, with the $\vmatch$ weighting, internal estimates that resample the data from a slicing into $\nsv$ sub-samples have the same statistical power as $\ns=\nsv$ independent simulation in determining the internal structure of the covariance matrix, matching the fact that both are drawn from Wishart distributions with the same degrees of freedom.

Eigenvectors corresponding to four different components for the real-space correlation function (highlighted with vertical arrows in Fig.~\ref{fig:eigenvalues_xir_baseline}) are shown in Fig.~\ref{fig:eigenvectors_xir_baseline}. Internal estimates of the eigenvectors corresponding to the main components agree very well with the reference ones. The level of agreement decreases with increasing eigennumber. However, internal methods provide reliable results even for the less important components if compared to estimates obtained using a number of independent simulations equal to the number of sub-samples $\nsv$ the original data are slice into. Estimates of the eigenvectors corresponding to the components with higher eigennumber ($25<i<30$) are very noisy even when performed using a set on 1000 independent datasets as can be seen in the bottom panel of Fig.~\ref{fig:eigenvalues_xir_baseline}.

The corresponding eigenvalues and eigenvectors for the baseline analysis for the multipole moments in redshift space are shown in  Fig.~\ref{fig:eigenvalues_mps_baseline} and  Fig.~\ref{fig:eigenvectors_mps_baseline} respectively. Eigenvalues in Fig.~\ref{fig:eigenvalues_mps_baseline} show a similar trend to that seen in real-space in Fig.~\ref{fig:eigenvalues_xir_baseline}. The main difference here between real- and redshift-space is the higher dimension of the data vector for the multipole moments that inevitably degrades the estimate of the covariance matrix especially for its less important components. Our baseline internal method provides accurate estimates of the eigenvalues for components which jointly account for $\sim85\%$ of the total variance. This number, that quantifies the accuracy of the internal estimate of the covariance matrix, is inevitably lower for the multipole moments compared to that for the real-space correlation function ($\sim99\%$) because we use the same number of realisations to estimate the covariance matrix both for the real- and redshift-space measurements despite the size of the multipole covariance matrix is 9 times larger than that of the covariance matrix for the real-space correlation function. The discrepancy increases for higher eigennumbers and is found to be more than $2\sigma$ for eigennumbers larger than $\sim30$ (out of 90 components). In particular, the deviation from the reference reaches $\sim -50\sigma$ for the component with the highest eigennumber. Another common factor with the real-space results is that internal methods provide estimates of the eigenvalues that are statistically identical to those obtained from a number of independent simulations equal to the number of sub-samples $\nsv$ the data are split into.

We complete the baseline analysis by comparing the internal and reference estimates of the eigenvectors of the multipoles covariance matrix. Although our baseline Jackknife method provides reliable eigenvectors for the main components, the agreement with the reference estimates rapidly degrades with increasing eigennumber. Finally we highlight the result that, although internal estimates do not agree very well with the reference from 1000 independent simulations for eigenvectors with higher eigennumbers, they match the results that uses only $\ns=\nsv$ independent simulations. This means that more accurate internal estimates of the covariance matrix are possible by increasing the number of sub-samples the data are split into.

\subsection{Impact of the Rescaling Schemes}     \label{sec:results_weightings}

\begin{figure}
    	\centering
		\includegraphics[width=\columnwidth]{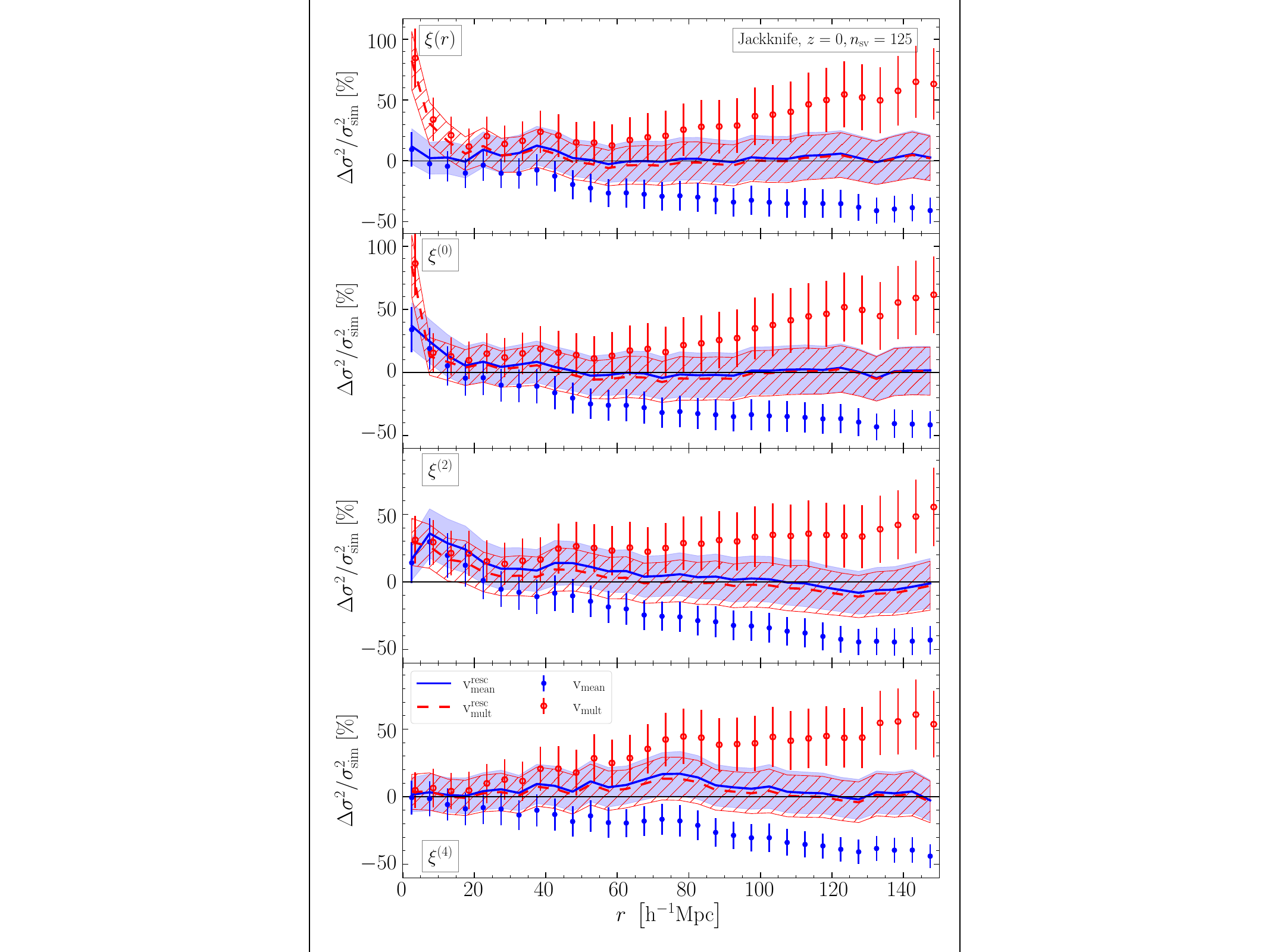}
		\caption{Comparison between raw and rescaled (using Eqns.~\eqref{eq:recsaling1} and~\eqref{eq:recsaling2}) versions of the $\vmean$- and $\vmult$-weighted Jackknife resampling method in recovering the input variances of the real-space two-point correlation function. Top panel: real-space correlation function. Bottom three panels: multipole moments of the redshift-space correlation function.}\label{fig:vars_xir_v0_resc}
\end{figure}

\begin{figure}
    	\centering
		\includegraphics[width=\columnwidth]{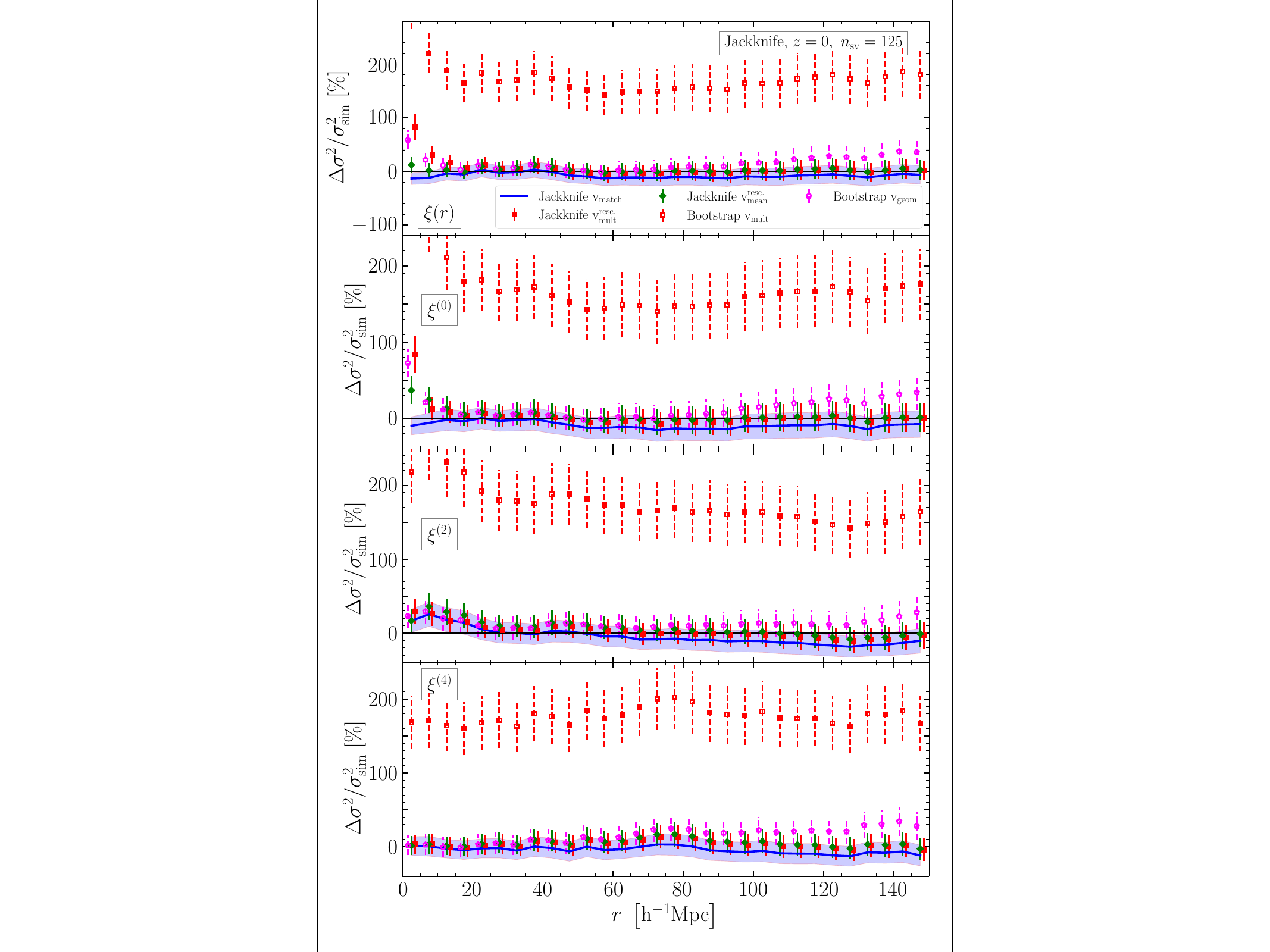}
		\caption{Relative difference between internal and reference estimates of the variances for the two-point correlation function. Top panel: real-space correlation function. Bottom three panels: multipole moments of the redshift-space correlation function. Blue thick lines with shaded bands shows the outcome of the baseline configuration for the Jackknife method (same as the blue lines in Fig.~\ref{fig:vars_xir_baseline}). For the Jackknife method we only show optimal estimates for each weighting scheme, i.e. applying the rescaling in Eqns.~\eqref{eq:recsaling1} and~\eqref{eq:recsaling2} to the $\vmean$ (green solid diamonds, same as the thick blue lines in Fig.~\ref{fig:vars_xir_v0_resc}) and $\vmult$ (red solid squares, same as the red dashed lines in Fig.~\ref{fig:vars_xir_v0_resc}) weightings, respectively. Results from the Bootstrap resampling using $\vmult$ and $\vgeom$ weighting scheme are shown as empty red squares and empty magenta pentagons with dashed error bars, respectively. For clarity of the plot we do not show here the Bootstrap estimates when using $\vmeanresc$ weights and report them in a separate plot in Fig.~\ref{fig:vars_xir_weighting_vmean}.} \label{fig:vars_xir_weighting}
\end{figure}

\begin{figure}
    	\centering
		\includegraphics[width=\columnwidth]{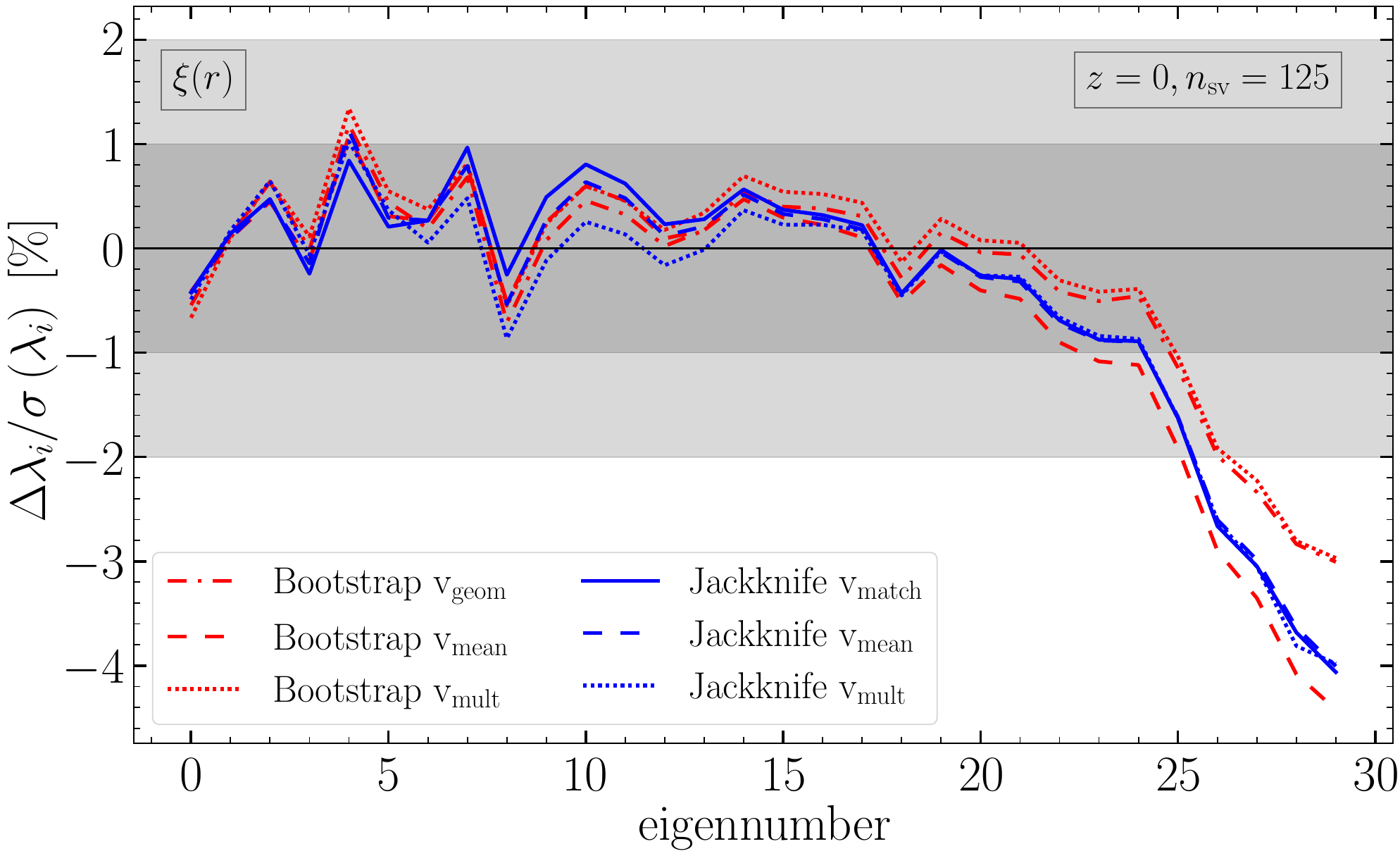}
		\caption{Same as in the bottom panel of Fig.~\ref{fig:eigenvalues_xir_baseline} but here we show results from all weighting schemes. The rescaling does not affect the correlation matrix and for this reason we only show the results from the raw weighting schemes. Since different estimates are very similar we only show their accuracy compared to the reference and omit the equivalent of the top panel in Fig.~\ref{fig:eigenvalues_xir_baseline}.}\label{fig:eigenvalues_xir_weighting}
\end{figure}

\begin{figure}
    	\centering
		\includegraphics[width=\columnwidth]{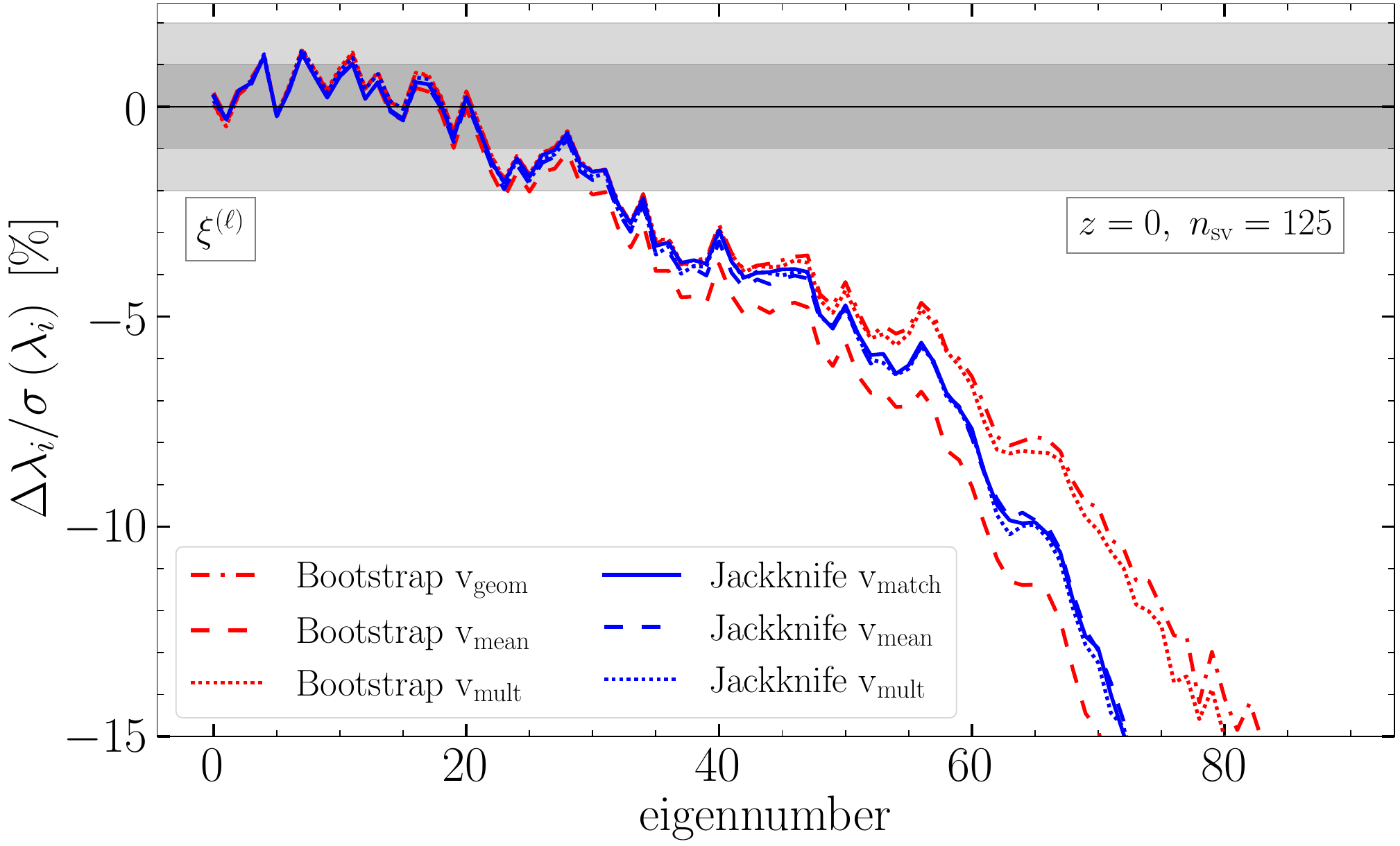}
		\caption{Same as in Fig.~\ref{fig:eigenvalues_xir_weighting}, here for the multipole moments of the redshift-space two-point correlation function.}\label{fig:eigenvalues_mps_weighting}
\end{figure}

In Fig.~\ref{fig:vars_xir_baseline} we compared the Jackknife variance estimates using the raw $\vmean$, $\vmult$ and $\vmatch$ weighting. In this section we analyse the improvements allowed by the rescaling introduced in Eqns.~\eqref{eq:recsaling1} and~\eqref{eq:recsaling2} for the $\vmean$- and $\vmult$-weighted Jackknife estimates of the variances. These factors are designed to correct for the mismatch in the cross-pairs count between Jackknife realisations and the reference estimates. 

The results plotted in the top panel of Fig.~\ref{fig:vars_xir_v0_resc} for the real-space correlation function and in the bottom three panels for the multipoles in redshift-space show that the rescaling is able to correct the offsets seen in Fig.~\ref{fig:vars_xir_baseline} for both the $\vmult$ and $\vmatch$ weighting schemes, bringing them into agreement with the baseline covariance. The effect of the rescaling is larger at large scales where the contribution of the cross pairs to the total pair count becomes non negligible. 

On small scales, we see an increase in the Jackknife variance recovered for the real-space and redshift-space monopole correlation functions for the $\vmult$ weighting, which is not corrected by the rescaling. We know that the origin of this must be linked to cross-pairs given that the strength of this effect changes for different weighting schemes, and they only differ in their effect on cross-pairs. The $\vmult$ weighting in a Jackknife analysis increases the recovered variance thanks to an increase in the contribution from the cross-pairs (see Section~\ref{sec:rescaling}), and this reveals the largest offset on small-scales. So the most likely explanation is that there is non-linear correlation between the small-scale pairs across a fixed boundary between subvolumes, such that the number of pairs weighted in each Jackknife realisation has a larger variance than that we would see comparing pairs with similar separations in different volumes as in our baseline covariance.

\subsection{Comparison between Jackknife and Bootstrap}
\label{sec:bootstrap}

We compare Jackknife and Bootstrap methods using different weighting schemes in Fig.~\ref{fig:vars_xir_weighting} for the real-space two-point correlation function (top panel) and for the multipole moments of the redshift-space correlation function (bottom three panels), respectively. For the Jackknife method we only show the optimal estimates we obtain after rescaling the $\vmean$- and $\vmult$-weighted resamplings using Eqns.~\eqref{eq:recsaling1} and~\eqref{eq:recsaling2}, respectively. For the $\vgeom$ weighting we will show only results from the Bootstrap resampling since for the Jackknife resamplings $\vgeom$ weighting is identical to the $\vmult$ weighting scheme. Results obtained using the rescaled $\vmean$-weighted Bootstrap method overlap very well with their Jackknife counterparts (see discussion in Section~\ref{sec:weighting}). This equality holds both in real- and redshift space and is confirmed by our numerical tests. We thus show the Bootstrap-Jackknife comparison for $\vmeanresc$ weighting in a separate plot in Fig.~\ref{fig:vars_xir_weighting_vmean} for the real-space quantities only. The relative performance of the different resampling configurations is very similar between real- and redshift space. In particular, the $\vmult$-weighted Bootstrap resamplings return a highly biased estimates of the variances, with an overestimate that is consistently larger than $\sim150\%$ at all scales. This is qualitatively consistent with the discussion in Section~\ref{sec:weighting} that $\vmult$-weighted Bootstrap method fails to accurately match terms of the covariance matrix that depend on the auto- and cross- pairs. The way that the Bootstrap method weights subvolumes then exacerbates this problem. The $\vgeom$-weighted Bootstrap is less aggressive, and provides more accurate estimates of the variance compared to the $\vmult$ weighting but it still overestimates the reference variances with a systematic offset that increases at larger pair separations where the contribution from the cross pairs becomes significant. All versions of the Jackknife resamplings shown in Fig.~\ref{fig:vars_xir_weighting} are unbiased given the error bars and agree with each other within less than $1\sigma$. We do not interpret deviations slightly larger than $1\sigma$ in a small number of bins as statistically meaningful.

The impact of different weightings on the eigenvalues of the covariance matrix is shown in Fig.~\ref{fig:eigenvalues_xir_weighting} and Fig.~\ref{fig:eigenvalues_mps_weighting} for the real-space correlation function and for the multipole moments of the redshift-space two-point correlation function, respectively. All resampling configurations yield very similar results to those seen in the baseline analysis for the main components. The $\vmult$- and $\vmean$-weighted Bootstrap resamplings provide marginally better results for the noisier components indicating a higher statistical power of the corresponding realisations. However, most of this improvement is in the regime where internal estimates deviate more than $2\sigma$ from the reference and as such are not reliable.

All resampling configurations provide estimates of the eigenvectors indistinguishable from each other. We thus do not show these results here and refer the reader to Fig.~\ref{fig:eigenvectors_xir_baseline} and Fig.~\ref{fig:eigenvectors_mps_baseline}. Jointly analysing the results for the eigenvalues and eigenvectors we conclude that all weighting schemes are able to capture the main structure of the covariance matrix given by its principal components.

\subsection{Impact of the Data Slicing}\label{sec:slicing}

\begin{figure}
    	\centering
		\includegraphics[width=\columnwidth]{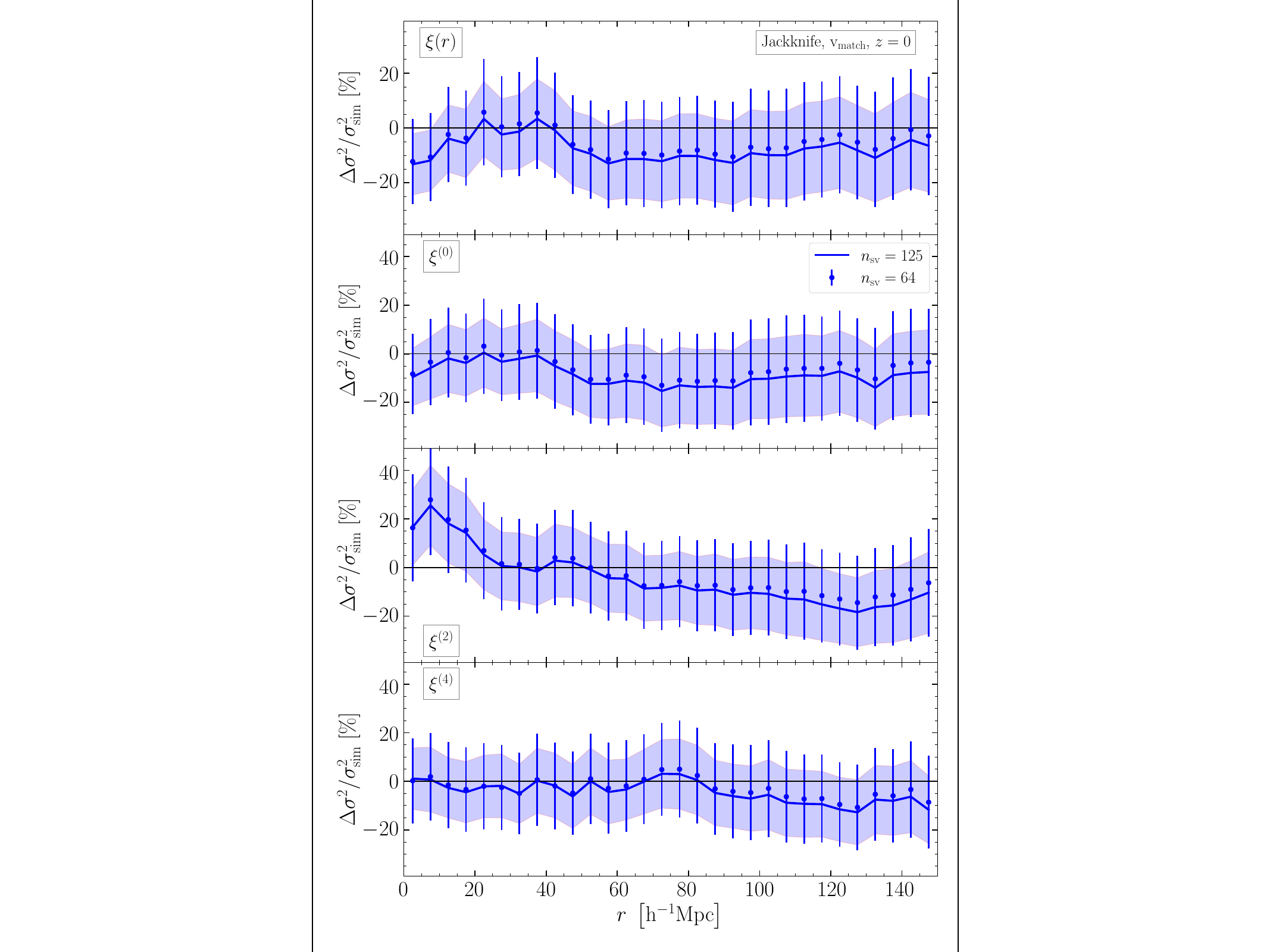}
		\caption{Accuracy of the $\vmatch$-weighted Jackknife estimates of the variances for the two-point correlation function at redshift $z=0$. Blue thick lines with shaded band show the baseline results (identical to those in Fig.~\ref{fig:vars_xir_baseline}) that use a slicing of the data into $\nsv=125$ sub-samples. Blue solid dots with errorbars are the same as the baseline results but split the data into a smaller number of sub-samples $\nsv=64$. Top panel: real-space correlation function. Bottom three panels: multipole moments of the redshift-space correlation function.}\label{fig:vars_xir_slicing}
\end{figure}

\begin{figure}
    	\centering
		\includegraphics[width=\columnwidth]{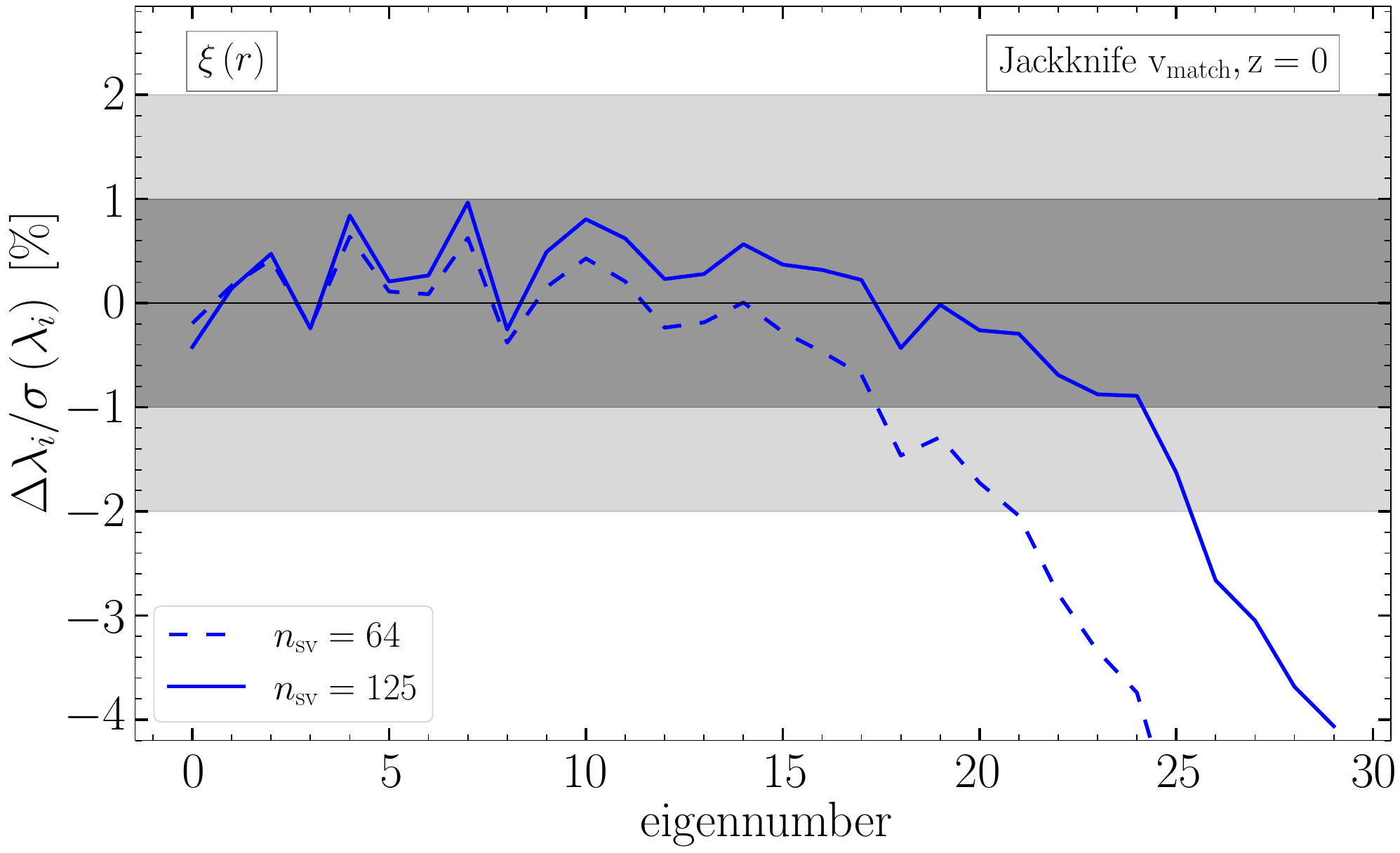}
		\caption{Difference, in units of $1\sigma$ standard deviation, between internal and reference estimates of the eigenvalues of the covariance matrix for the real-space two-point correlation function at redshift $z=0$. Blue thick line is the same as in the bottom panel of Fig.~\ref{fig:eigenvalues_xir_baseline} for the baseline configuration with $\nsv=125$ sub-samples slicing. Blue dashed line is the same as the baseline result but with a slicing of the data into $\nsv=64$ sub-samples.}\label{fig:eigenvalues_xir_slicing}
\end{figure}

\begin{figure}
    	\centering
		\includegraphics[width=\columnwidth]{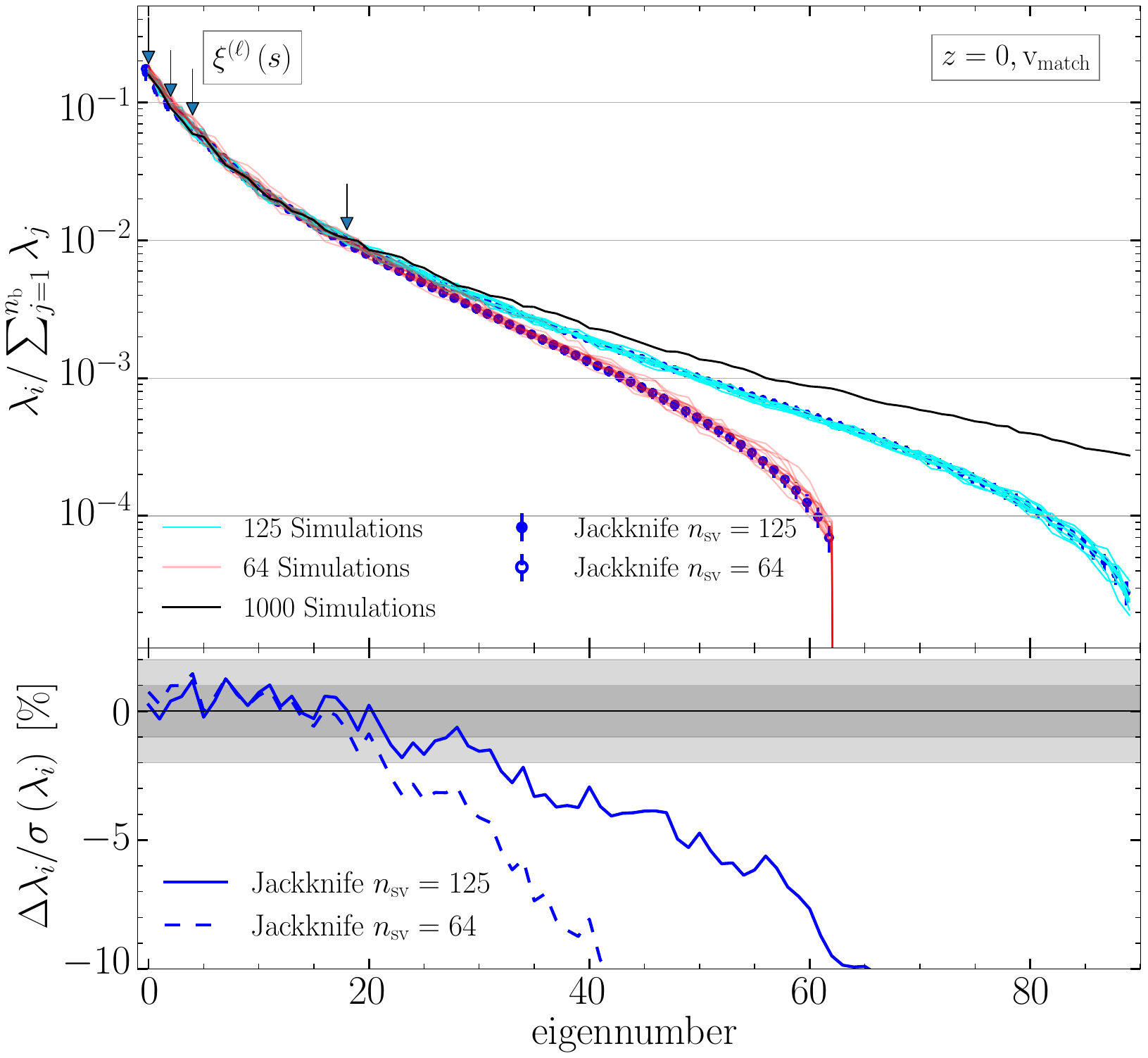}
		\caption{Same as in Fig.~\ref{fig:eigenvalues_mps_baseline} but here we compare the $\vmatch$-weighted Jackknife estimates of the eigenvalues for two different data slicings into $\nsv=125$ (filled blue dots in the top panel and thick line in the bottom panel) and $\nsv=64$ (empty blue dots in the top panel and dashed line in the bottom panel) subvolumes at $z=0$. With respect to the Fig.~\ref{fig:eigenvalues_xir_slicing} for the results in real space, here we also show the eigenvalues in the top panel to highlight the sudden drop at eigenxnumber = 63 that indicates a singular covariance matrix.}\label{fig:eigenvalues_mps_slicing}
\end{figure}

We now investigate how changing the number of sub-samples, the data are split into, from $\nsv=125$ to $\nsv=64$ alters the baseline results presented in Section~\ref{sec:results_baseline}. To keep the comparison straightforward we use the same data at redshift $z=0$ and change only the number of sub-samples $\nsv$.

The results for the estimates of the variances are shown in Fig.~\ref{fig:vars_xir_slicing} for the real-space two-point correlation function (top panel) and the multipole moments of the redshift-space two-point correlation function (bottom three panels). In both real- and redshift space we find that the $\vmatch$ weighting is able to accurately predict the variances at all scales regardless of the number of sub-samples $\nsv$ used to slice the data. Some differences between the two values of $\nsv$ are found at the largest scales explored here but they are well below the level of statistical uncertainties. Slicing the data into $\nsv=125$ sub-samples is found to provide marginally improved errorbars on the variance estimates compare to the case of $\nsv=64$.

Results for the eigenvalues are shown in Fig.~\ref{fig:eigenvalues_xir_slicing} and Fig.~\ref{fig:eigenvalues_mps_slicing} for the real- and redshift-space quantities. Increasing the number of sub-sampled used to slice the data is found to provide a major improvement in the agreement between internal and reference estimates of the eigenvalues. Although these improvement may not seem large enough for the cases tested in this work, the number of sub-samples $\nsv$ is the main variable among those explored here that provides the major boost in the accuracy of the internal estimates of the correlation matrix that ultimately captures the off-diagonal structure of the corresponding data covariance matrix. One important point to notice here is that, in order to obtain non-singular estimates of the covariance matrix using the baseline $\vmatch$-weighted Jackknife method, the number of sub-samples $\nsv$ must be larger than the number of data points. Indeed, in the case of the multipole moments the size of the data vector is $90$ and consequently the $\nsv=64$ slicing yields a singular covariance matrix as shown by the sudden drop in the eigenvalues in the top panel of Fig.~\ref{fig:eigenvalues_mps_slicing} (red thin lines and empty blue dots).

Apart from a minor degradation in the the statistical errors, the eigenvectors estimated using $\nsv=64$ sub-samples slicing do not show any detectable difference with respect to those shown in Fig.~\ref{fig:eigenvectors_xir_baseline} and~\ref{fig:eigenvectors_mps_baseline} for the $\nsv=125$ case. We thus do not show the corresponding plots here and refer the reader to Fig.~\ref{fig:eigenvectors_xir_baseline} and Fig.~\ref{fig:eigenvectors_mps_baseline}.

We find very similar results for the $\vmultresc$- and $\vmeanresc$-weighted Jackknife estimates of the covariance matrix for both real-space correlation function and for the multipole moments in redshift space.

\subsection{Robustness Against Intrinsic Clustering}\label{sec:clustering}

\begin{figure}
    	\centering
		\includegraphics[width=\columnwidth]{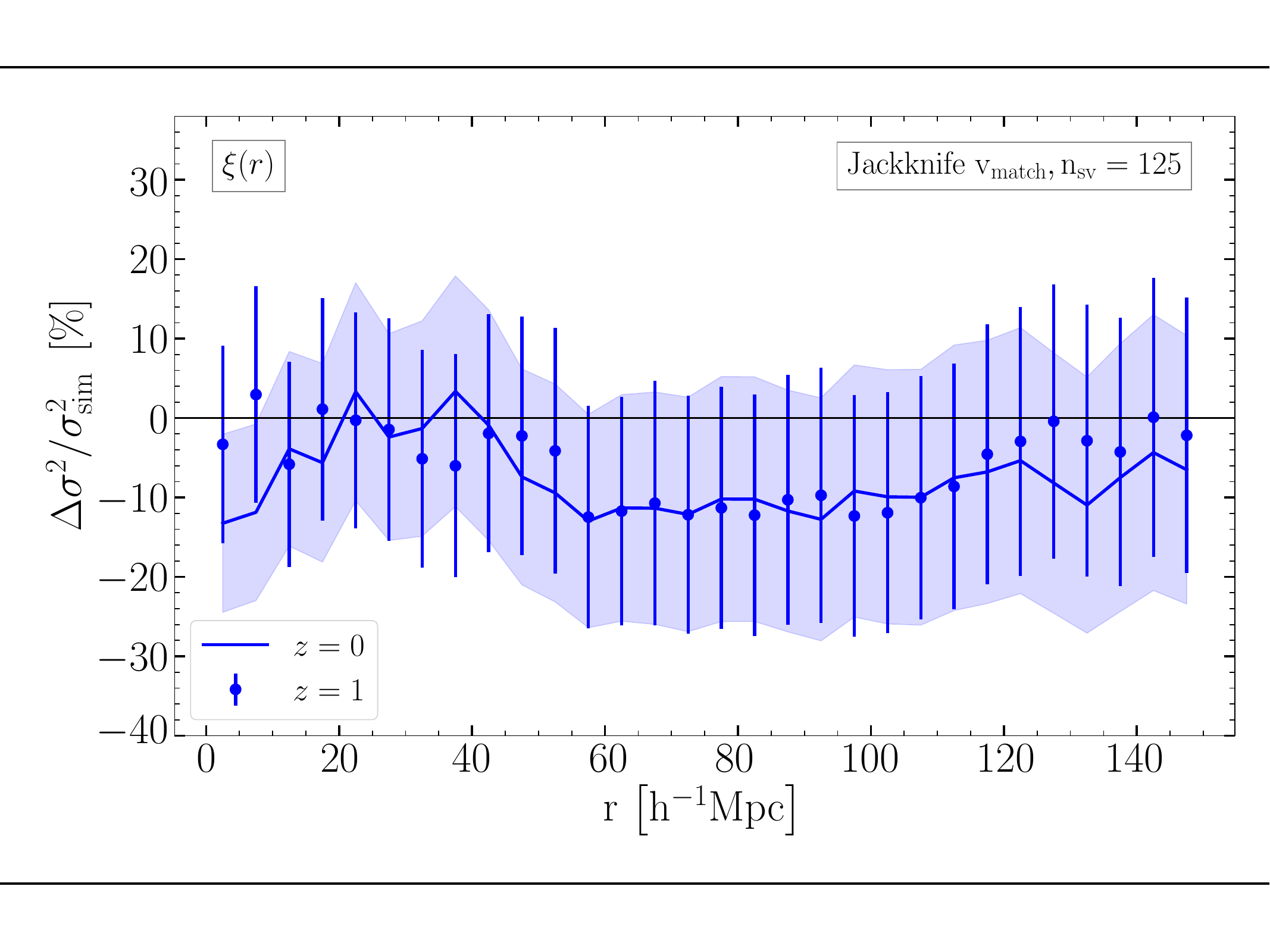}
		\caption{Comparison between the accuracy of the $\vmatch$-weighted Jackknife at two different redshifts of $z=0$ (blue thick line with shaded band, same as the blue thick line in the top panel of Fig.~\ref{fig:vars_xir_baseline}) and $z=1$ (blue dots with error bras). Here we show only the results for the real-space two-point correlation function. The outcome of these tests for the multipoles in redshift space show a very similar trend.}\label{fig:vars_xir_baseline_zs1}
\end{figure}

\begin{figure}
    	\centering
		\includegraphics[width=\columnwidth]{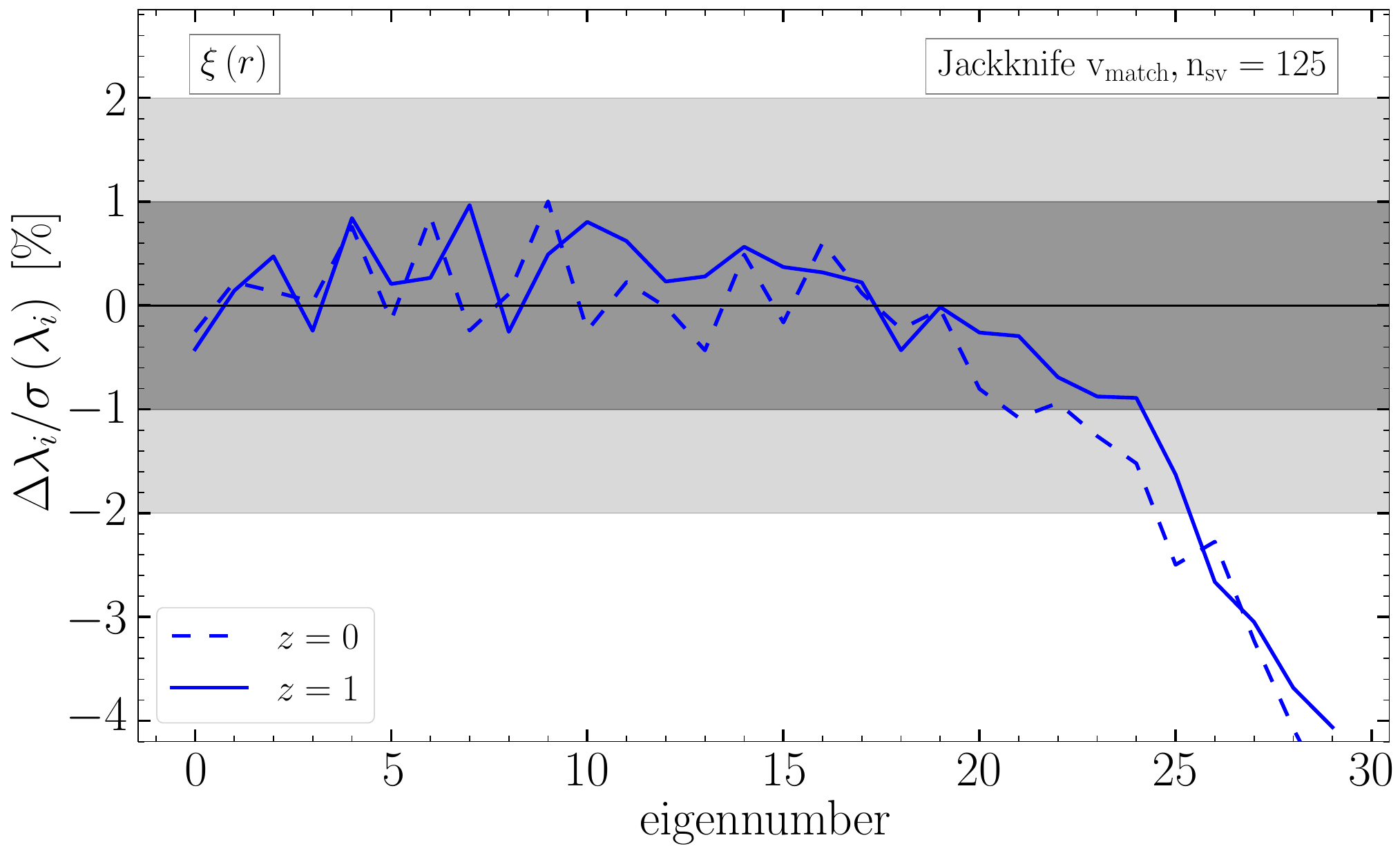}
		\caption{Deviations, in units of $1\sigma$ statistical errors, of the $\vmatch$-weighted Jackknife estimates of the eigenvalues with respect to the reference ones. Here we compare the results at two redshifts of $z=0$ (blue thick line, same as in the bottom panel of Fig.~\ref{fig:eigenvalues_xir_baseline}) and $z=1$ (blue dashed line). We show results for the covariance matrix of the real-space correlation function. Results for the multipoles in redshift space show a very similar trend.}\label{fig:eigenvalues_xir_baseline_zs1}
\end{figure}

So far we have tested the dependence of the method on different properties of the resampling techniques such as the weighting schemes, rescaling and data slicing. We perform one last robustness test to check the reliability of our results against the intrinsic clustering of the sample being used. For this purpose we repeat the baseline analysis of Section~\ref{sec:results_baseline} but using simulated catalogues of dark matter halos at a redshift of $z=1$. These samples, although built using the same cosmological model adopted for the ones used in the baseline analysis, exhibit a significantly different intrinsic clustering as shown in Fig.~\ref{fig:xir_sims}. For the sake of clarity in this section we show only the results for the real-space two-point correlation function following the same structure of Section~\ref{sec:results_baseline}. We also limit our comparison to the variances and eigenvalues only. Eigenvectors are found to be marginally affected by the change in clustering. The results are shown in Fig.~\ref{fig:vars_xir_baseline_zs1} for the variances and in Fig.~\ref{fig:eigenvalues_xir_baseline_zs1} for the eigenvalues. Despite the difference in clustering, results at redshift $z=1$ are virtually identical to those from the baseline analysis at redshift $z=0$. We find the same trend also for the covariance matrix of the multipole moments of the redshift-space two-point correlation function. This applies to all three elements of the covariance matrix, namely the variances, eigenvalues and eigenvectors. The same trend is observed for the $\vmultresc$- and $\vmeanresc$-weighted Jackknife estimates of the covariance matrix. We thus conclude that the improvements proposed for the Jackknife method in this work generalise very well to samples different than those used here.

\section{Summary and Conclusions}

In this work we have analysed the performance of the Bootstrap and the delete-one Jackknife internal estimates of the data covariance matrix for the two-point correlation function. Internal estimates split the observed data into $\nsv$ sub-samples and use different resampling techniques to make multiple copies of the observations. While for the one-point statistics we have a unique choice to assign the weight to each object in the resampled data, there is an ambiguity in how to define weights for pairs of objects, the building blocks of the two-point correlation function estimators. We test four different weighting schemes: the default $\vmult$ weighting; the $\vmean$ weighting designed to match pair numbers; the $\vgeom$ weighting proposed to provide a better estimate of the cosmic-variance term and the $\vmatch$ weighting that matches the weighting of auto- and cross-pairs between sub-samples in the Jackknife methodology. 

We use a set of 1000 independent QUIJOTE simulations both to estimate the reference covariance matrix and to perform the internal estimates. In turn, for each internal method we obtain 1000 estimates of the data covariance matrix, each using a single simulation. We infer the accuracy of each internal method as the mean difference between the related 1000 estimates of the covariance matrix and the reference one while the corresponding standard deviation is used as the measure for the statistical precision expected for a single dataset. While we would have liked to include results from a larger sample of simulations, our analysis was limited by the computational resources and time available. As Fig.~\ref{fig:vars_xir_baseline} shows, this sample is sufficient to clearly distinguish between results from different weighting schemes, and demonstrate that the $\vmatch$ weighting scheme provides less biased results than more traditional methods, the primary result of our work.

We perform our analysis using the isotropic real-space two-point correlation function and the multipole moments of the anisotropic redshift-space two-point correlation function. We first decompose the covariance matrix into its diagonal elements, the variances, and the correlation matrix (also referred to as the normalised covariance matrix). The correlation matrix is then analysed through the `principal component analysis' (pca) in terms of its eigenvalues and eigenvectors. The comparison between two estimates of the covariance matrix is carried out by comparing the related vectors of variances and the eigenvalues and eigenvectors of the corresponding correlation matrix.

We find that, both in real- and redshift space, the $\vmatch$-weighted Jackknife method provides accurate estimates of the variance at scales up to $150\mhmpc$ considered in this work. In comparison, the raw $\vmean$ weighting recovers the reference estimates within $1\sigma$ but only at the shot-noise dominated small scales ($\lesssim 30 \mhmpc$) and underestimates them at larger scales. The raw $\vmult$ weighting scheme overestimates the reference values of the variance over all scales. Both raw $\vmean$ and $\vmult$-weighted Jackknife methods fail to recover the reference variances at large scales due to a mismatch in the cross-pairs counts between Jackknife realisations and the original data. We propose, in Eqns.~\eqref{eq:recsaling1} and~\eqref{eq:recsaling2}, a simple correction for both $\vmean$ and $\vmult$-weighted Jackknife  realisations. The rescaling allows us to successfully recover unbiased estimates of the variances at scales $\gtrsim 5\mhmpc$. These improvements hold both in real- and redshift space. 

We also analyse the performance of the Bootstrap resampling technique using similar weighting schemes. The $\vmult$-based Bootstrap estimates of the variances are found to be the most biased due to a mismatch in how auto- and cross-pairs are treated. These effects are exacerbated with respect to their Jackknife equivalent due to the `resampling with replacement' method and lead to an overestimate of the variance that is consistently higher than $\sim150\%$ ($\sim60\%$ for the standard deviations) over all scales up to $150\mhmpc$. The $\vmean$ weighted Bootstrap realisations made using the slicing of the data into $\nsv$ sub-samples can be viewed as a linear combination of the corresponding Jackknife copies of the same data. As such both Bootstrap and Jackknife estimates of the variances are expected to be very similar. This is confirmed by the numerical tests presented in Appendix~\ref{app:appendix_vmean}. The $\vgeom$ weighting scheme corrects some of the issues affecting the $\vmult$ weighting as it changes the cross-pair contribution, and therefore reduces the systematic offset with respect to the $\vmult$ weighting but still overestimate the variances by up to $\sim50\%$. The rescaling factor proposed in Eqns.~\eqref{eq:recsaling2} is less successful for the Bootstrap method, as it is not designed to correct for these effects.

We have analysed the off-diagonal structure of the covariance matrix, as captured by the correlation matrix, by splitting into an elementwise comparison of the corresponding normalised eigenvalues and eigenvectors. The main conclusion of this principal component analysis is that different weighting schemes, either applied to Jackknife or Bootstrap resampling methods, provide estimates of the eigenvectors indistinguishable from each other given the statistical errors. All internal methods explored in this work were able to recover the reference eigenvalues for components that together account for $\gtrsim85\%$ of the total variance ($>85\%$ for the multipole moments and $>99\%$ for the real-space correlation function) but fail to do so for the noisier components. For a fixed number of sub-samples $\nsv$, used to slice the data, all variants of the Jackknife provide statistically identical estimates of the eigenvalues. Furthermore, although the eigenvalues resulting from the Jackknife method do not match the reference ones obtained from the set of 1000 independent simulations, they are found to be in a very good agreement with those obtained using a number of simulations $\ns$ equal to the number of sub-samples $\nsv$ the data are sliced into. This is expected, given that both can be considered to be drawn from a Wishart distribution with the same degrees of freedom. This indicates that in order to obtain a Jackknife estimate of the covariance matrix that matches in accuracy the one obtained from a set of $\ns$ independent simulations, we need to use $\nsv=\ns$ independent Jackknife realisations, i.e. slice the data into $\nsv=\ns$ sub-samples. Obviously, if the number of sub-samples is smaller than the number of bins in the measurement vector $\nsv\le \nb$ the Jackknife-estimated covariance matrix will be singular and thus non-invertible. As mentioned previously, we expect both the Bootstrap and Jackknife methods to provide very close estimates of the covariance matrix when carried out using the $\vmean$ weighting scheme. This is seen for the variances in Fig.~\ref{fig:vars_xir_weighting_vmean} and confirmed with the pca in Fig.~\ref{fig:eigenvalues_xir_weighting} and  Fig.~\ref{fig:eigenvalues_mps_weighting}.

We have also performed a number of robustness tests of the Jackknife method to test the reliability of the improvements proposed in this work. In particular, we used datasets at redshifts of $z=0$ and $z=1$ to test the dependence of the results on the intrinsic clustering of the sample being analysed. We find no statistically meaningful difference in the results at the two redshifts. This, and the fact that our weighting schemes are general in nature and not tied to any property of our simulations, suggests that it will be straightforward to generalise our results to other datasets based on either a different cosmological model or a different kind of tracer. We also analyse how a change in the slicing of the original data into different number of sub-samples affects our conclusions. We test two different slicings into $\nsv=64$ and $\nsv=125$ sub-samples. Our results on the estimates of the variances are robust against these two slicings of the data. We find a major change in the recovered eigenvalues as discussed above. In particular, using a number of sub-samples $\nsv=125$, i.e. closer to the number of simulations $\ns=1000$ used for the reference estimates, improves the agreement between the internal estimates and the reference ones. This naturally arises from the fact that a larger number of sub-samples allows us to draw a larger number of independent Jackknife realisations.

To summarise, we find that the Jackknife resampling technique can be adjusted to provide reliable estimates of the covariance matrix for the two-point correlation function. We have proposed three different ways to do so by means of the $\vmeanresc$, $\vmultresc$ and $\vmatch$ weighting schemes. Although $\vmatch$ is more consistent with a statistical derivation in that it provides a single weighting for pairs that requires no further adjustment, we find that $\vmeanresc$ provides the least biased estimates of the variances and thus of the covariance matrix at all scales. However, it is not clear whether this is due some systematic effect not accounted for by our method or a simple statistical noise. Furthermore, all of these three methods utilise the same underlying theory about the causes of the discrepancies, and so there is no theoretical reason to choose between them, except that $\vmatch$ makes slightly fewer assumptions as no rescaling is required.

The modifications to the basic Jackknife resampling method proposed in this work should provide a major improvement in the accuracy of the technique in recovering the reference data covariance matrix. Although we see a large improvement over the accuracy of the standard Jackknife method, our proposed method is obviously limited by a number of factors: in particular, we have proposed corrections based on pair numbers, and this does not necessarily correct for sample variance. It is difficult to correct for both shot-noise and sample variance as cross-pairs mean that any internal method for calculating the covariance will struggle to slice the volume in a way that allows any method to correctly takes into all contributions to the covariance.

Our comparison against the baseline has also been constrained by the number of runs that we have completed, and could be improved in a number of ways: Firstly, a larger number of independent simulation than that used here (1000) would be helpful in highlighting any systematic beyond the``statistical resolution'' of our analysis. Secondly, in this work we tested only two different slicing ($\nsv$) of the data. Extending these tests to larger values of $\nsv$ would provide the optimal slicing required to accurately predict the structure of the data covariance matrix. Nevertheless, our analysis is sufficient that the broad trends are clear and match theoretical expectation, suggesting that our proposed method can be considered a first order correction to this problem.

\section*{Acknowledgements}

Research at Perimeter Institute is supported in part by the Government of Canada through the Department of Innovation, Science and Economic Development Canada and by the Province of Ontario through the Ministry of Colleges and Universities. FGM \& WJP acknowledge support from the Canadian Space Agency (CSA) and the Natural Sciences and Engineering Research Council of Canada (NSERC).

This research was enabled in part by support provided by Compute Ontario (www.computeontario.ca) and Compute Canada (www.computecanada.ca).

\section*{Data Availability}

The simulations used in this work are publicly available at: https://quijote-simulations.readthedocs.io



\bibliographystyle{mnras}
\bibliography{bibliography} 


\appendix

\section{$\vmeanresc$-weighted Boostrap and Jackknife}\label{app:appendix_vmean}

\begin{figure}
    	\centering
		\includegraphics[width=\columnwidth]{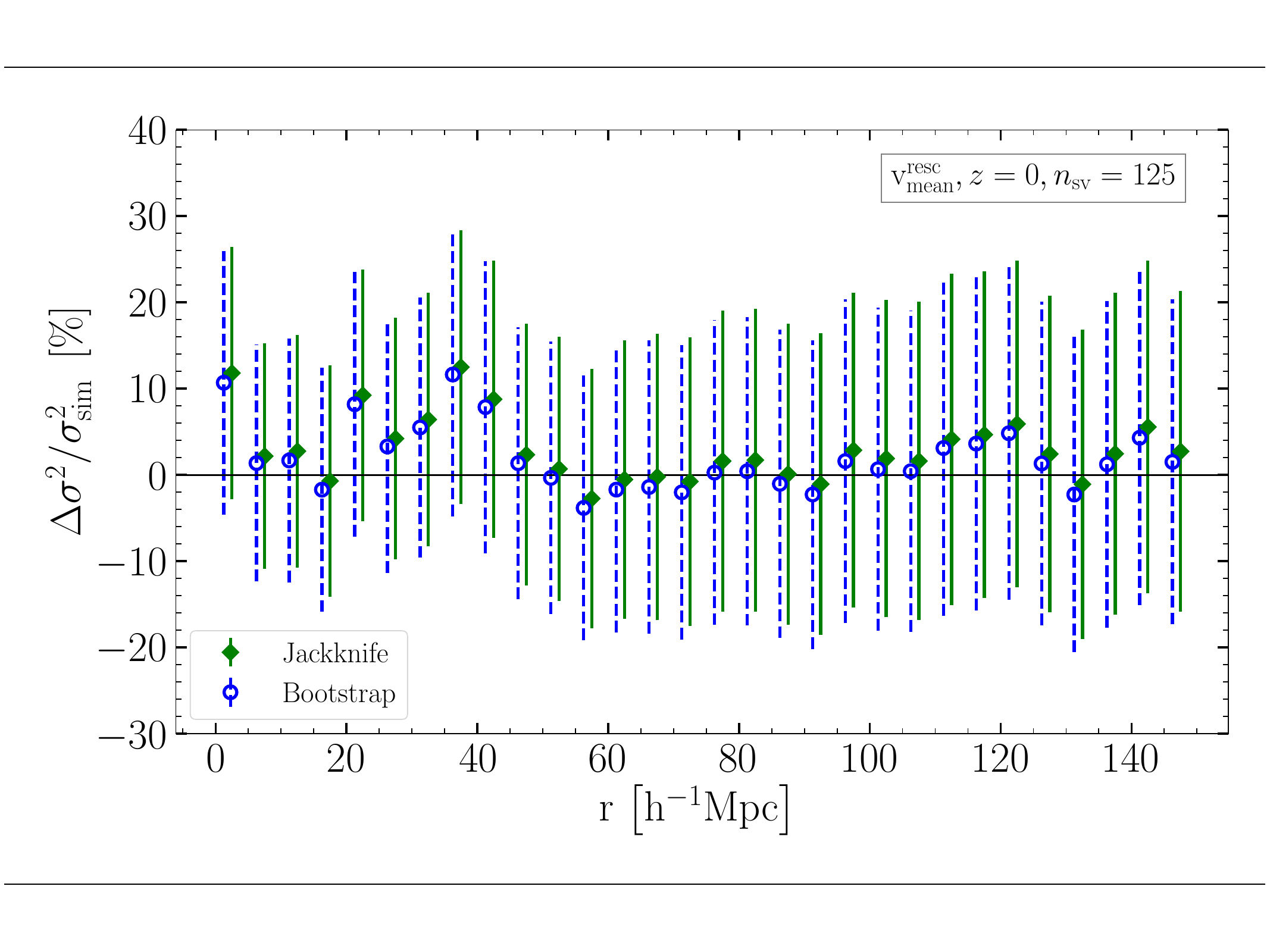}
		\caption{Comparison between the accuracy of the $\vmeanresc$-weighed Jackknife and Bootstrap resampling methods in recovering the input variances of the real-space two-point correlation function. For comparison green diamonds are the same as in Fig.~\ref{fig:vars_xir_weighting}.} \label{fig:vars_xir_weighting_vmean}
\end{figure}

In Sec.~\ref{sec:weighting} we have discussed the equality between the Bootstrap and Jackknife resamplings when performed using the $\vmean$ weighting. This was shown to be valid for the eigenvalues and eigenvectors of the correlation matrix in Sec. \ref{sec:results_weightings}. In Fig.~\ref{fig:vars_xir_weighting_vmean} we show that Jackknife and Bootstrap estimates of the variances for the real-space two-point correlation function at $z=0$ performed using the $\vmean$ weighting, and rescaled using Eqns.~\eqref{eq:recsaling1}, indeed match each other very closely over all scales. Although not shown here, we find identical results for the multipole moments of the redshift-space correlation function.

\bsp	
\label{lastpage}
\end{document}